\documentclass[times]{article}

\usepackage{moreverb}

\usepackage[colorlinks,bookmarksopen,bookmarksnumbered,citecolor=red,urlcolor=red]{hyperref}
\usepackage{mathrsfs}
\usepackage{amsfonts}
\usepackage{amssymb}
\usepackage{stmaryrd}
\usepackage{graphicx}
\usepackage{amsbsy,amsmath}

\usepackage[ruled,slide]{algorithm2e}

\newcommand\BibTeX{{\rmfamily B\kern-.05em \textsc{i\kern-.025em b}\kern-.08em
T\kern-.1667em\lower.7ex\hbox{E}\kern-.125emX}}

\DeclareMathOperator*{\A}{ \raisebox{-2pt}{$\mathlarger{\mathlarger{\mathlarger{\mathlarger{\mathlarger{\mathsf{A}}}}}}$} }

\begin{document}

%\runningheads{A.~N.~Other}{A demonstration of the \journalabb\
%class file}

%\title{Arc-Length control for brittle material with configurational-force-driven crack propagation
%\footnotemark[2]}

%\title{Thermodynamically consistent framework for three--dimensional brittle
%fracture with configurational-force-driven crack propagation \footnotemark[2]}

\title{Three--dimensional brittle
fracture:\\ configurational-force-driven crack propagation}

\author{
\L ukasz~Kaczmarczyk, Mohaddeseh Mousavi Nezhad, Chris Pearce\\
School of Engineering, University of Glasgow, G12 8QQ, UK\\
e-mail: Lukasz.Kaczmarczyk@glasow.ac.uk, Chris.Pearce@glasgow.ac.uk
}

\maketitle

\begin{abstract}

This paper presents a computational framework for quasi-static brittle fracture
in three dimensional solids. The paper set outs the theoretical basis for
determining the initiation and direction of propagating cracks based on the
concept of configurational mechanics, consistent with Griffith's theory.
Resolution of the propagating crack by the finite element mesh is achieved by
restricting cracks to element faces and adapting the mesh to align it with the
predicted crack direction. A local mesh improvement procedure is developed to
maximise mesh quality in order to improve both accuracy and solution robustness
and to remove the influence of the initial mesh on the direction of propagating
cracks. An arc-length control technique is derived to enable the dissipative
load path to be traced. A hierarchical hp-refinement strategy is implemented in
order to improve both the approximation of displacements and crack geometry.
The performance of this modelling approach is demonstrated on two numerical
examples that qualitatively illustrate its ability to predict complex crack
paths. All problems are three-dimensional, including a torsion problem that
results in the accurate prediction of a doubly-curved crack.   

\end{abstract}

%\footnotetext[2]{Please ensure that you use the most up to date
%class file,
%available from the NME Home Page at\\
%\href{http://www3.interscience.wiley.com/journal/1430/home}{\texttt{http://www3.interscience.wiley.com/journal/1430/home}}}

\section{Introduction}

Fracture is a pervasive problem in materials and structures and predictive
modelling of crack propagation remains one of the most significant challenges
in solid mechanics. A computational framework for modelling crack propagation
must not only be able to predict the initiation and the direction of cracks but
also to provide a numerical setting to accurately resolve the crack path.

In the Finite Element Method, strategies for discretization of the discontinuities can
be categorized as either smeared or discrete. The former is
attractive from the point of view that the problem can be solved within a
continuum setting, without the need for approximation of discontinuities or
changing mesh connectivity. However, as strains localize, numerical difficulties arise and regularization is required. Discrete approaches are 
able to directly approximate macroscopic crack geometry and therefore describe fractures in a more natural and straightforward
manner in terms of displacement jumps and tractions. Developments include introducing embedded displacement jumps within finite
elements via additional enhanced strain modes (for example \cite{R1}) or with enrichment functions in the context of the partition of unity (PoU) method, e.g. \cite{R2}.

A straightforward discrete crack approach is to restrict the crack path to
element faces. In the simplest case, the predicted path can be
strongly influenced by the mesh, thereby influencing the crack surface area,
and strongly affecting the amount of energy dissipated. The crack path's
dependence on the mesh can be somewhat reduced by using very fine, unstructured
meshes, but this results in computationally expensive analyses. It is worth noting that the
authors \cite{R3,R4} have shown that within a cohesive crack methodology applied to heterogeneous
microstructures, the crack propagation is largely controlled by the need for
the mesh to resolve the heterogeneities. However, in the modelling of ideally
brittle homogeneous materials, studied here, this is clearly not the case. 

In addition to the need for the mesh to resolve the crack, it is necessary to employ some rational means of determining
if a crack will propagate and, if so, the direction of crack propagation.
The approach taken in this
paper is based on the principle of maximal energy dissipation, using
configurational forces to determine the direction of the propagating crack front.
A similar technique was successfully adopted by a number of authors, but here we
mainly follow the work of Miehe and colleagues~\cite{R5,R6}.
 
This paper is primarily concerned with predicting crack propagation in large
three-dimensional problems. The efficiency of such problems, with a large
number of degrees of freedom, usually requires the use of an iterative solver
for solving the system of algebraic equations. In this case, it is important to
control element quality in order to optimise the system matrix conditioning,
thereby increasing the computational efficiency of the solver. This can be
problematic in methods such as PoU, where the enrichment functions not only the
increases the bandwidth of the stiffness matrix but also the matrix
conditioning deteriorates \cite{R24}. In this paper we will show how
controlling the mesh quality improves both the crack path predictions and the
robustness of the solution algorithm. Local r-adaptivity is adopted to mitigate the influence of
the mesh.

Two numerical examples for crack propagation are presented that demonstrate the
ability of the formulation to accurately predict crack paths without bias from the underlying mesh and the 
influence of mesh adaptivity and mesh quality control on the solution obtained.

\section{Body and crack kinematics}

\begin{figure}[th]
\setlength{\fboxsep}{0pt}%
\setlength{\fboxrule}{0pt}%
\begin{center}
\includegraphics[width=0.7\textwidth]{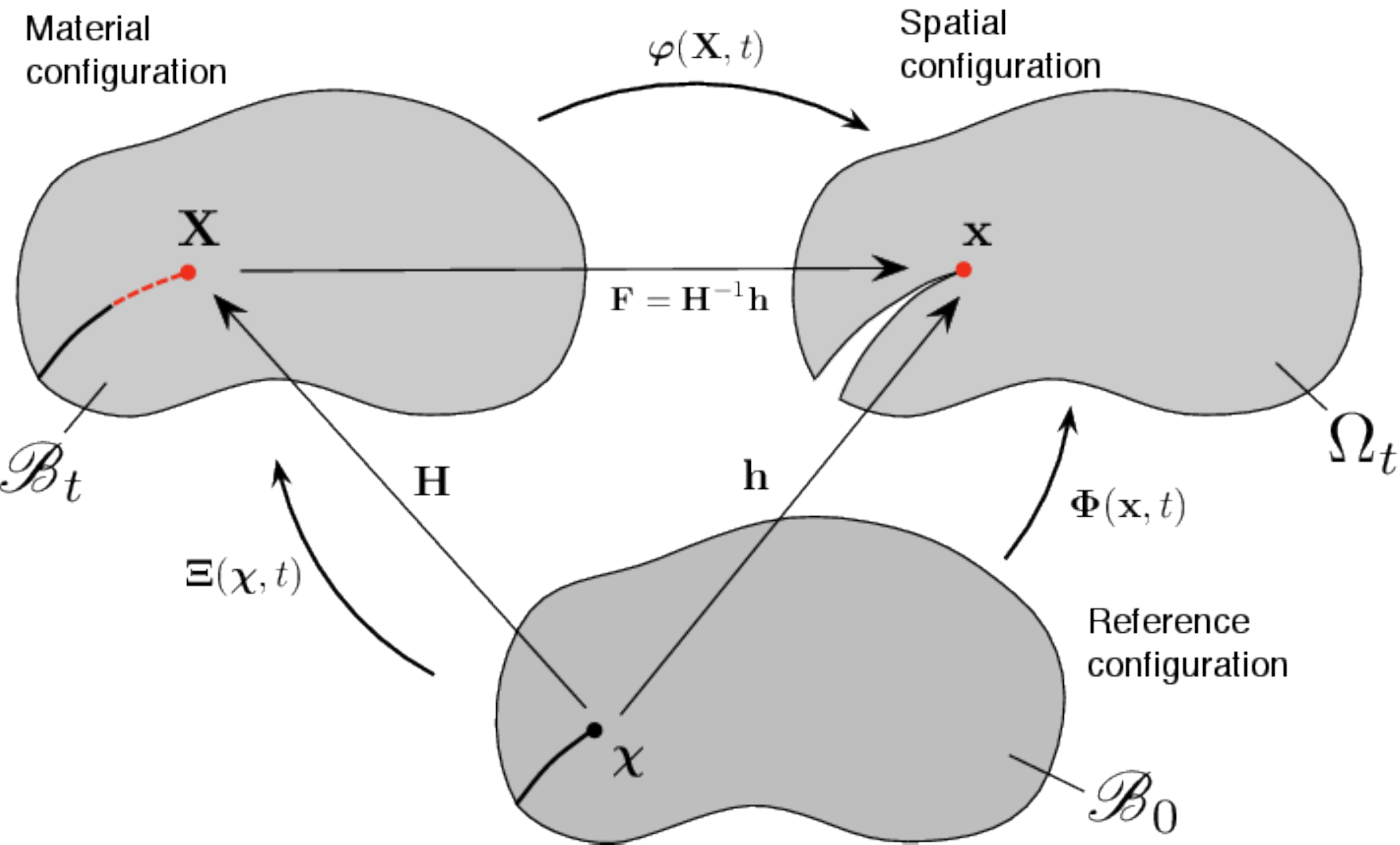}
\end{center}
\caption{Deformation and structural change of a body with a propagating crack.\label{F1}}
\end{figure}

Continuous evolution of propagating cracks are analysed within the context of the
Arbitrary Lagrangian Eulerian description. This is expressed by differentiable
mappings from the reference material configuration to both the current spatial
configuration and the current material configuration.
In the context of this paper, these mappings are utilized to independently observe the
deformation of material in the physical space $\Omega_t$ and the evolution of the crack
surface in the material space  $\mathscr{B}_t$, see Fig.~\ref{F1}.

The material coordinates
$\mathbf{X}$ are mapped onto the spatial coordinates $\mathbf{x}$ via the
familiar deformation map $\boldsymbol\varphi$, so that
\begin{equation}
{\boldsymbol\varphi}:\mathscr{B}_t \to \Omega_t,\,\mathbf{x}={\boldsymbol\varphi}(\mathbf{X},t)
\end{equation}
The mapping ${\boldsymbol\varphi}$ must be differentiable, injective, and
orientation preserving except at the boundary.
The physical displacements are:
\begin{equation}
\mathbf{u}=\mathbf{x}-\mathbf{X}
\end{equation}
The reference configuration describes the body before crack extension, with mapping ${\boldsymbol\Xi}$ of the reference coordinates $\boldsymbol\chi$ on to the material coordinates $\mathbf{X}$  as:
\begin{equation}
{\boldsymbol\Xi}:\mathscr{B}_0 \to \mathscr{B}_t,\,\mathbf{X}={\boldsymbol\Xi}(\boldsymbol\chi,t)
\end{equation}
This mapping represents a material structural change, which, in the context of this work, is an extension of the crack due to movement of the crack front. We can also define a mapping ${\boldsymbol\Phi}$ of the reference coordinates on to the spatial coordinates as:
\begin{equation}
{\boldsymbol\Phi}:\mathscr{B}_0 \to \Omega_t,\,\mathbf{x}={\boldsymbol\Phi}(\boldsymbol\chi,t)
\end{equation}

For convenience, we define a composite space-time vector containing both a
coordinate vector and time. For example:
\begin{equation}
(\boldsymbol\chi,t)=(\chi_1, \chi_2, \chi_3, t)
\end{equation}
Thus, the following derivatives can be expressed as
\begin{equation}
\label{eq::mat_map}
\frac{\partial \{{\boldsymbol\Xi}(\boldsymbol\chi,t),t\}  }
{\partial (\boldsymbol\chi,t)} 
= 
\left[
\begin{array}{cc}
\frac{\partial \boldsymbol{\Xi}}{\partial \boldsymbol\chi} &
\frac{\partial \boldsymbol{\Xi}}{\partial t} \\[0.5em]
\mathbf{0}^\textrm{T} & 1
\end{array}
\right] = 
\left[
\begin{array}{cc}
\mathbf{H} & \dot{\mathbf{W}} \\[0.5em]
\mathbf{0}^\textrm{T} & 1
\end{array}
\right]
\end{equation}
\begin{equation}
\frac{\partial \{{\boldsymbol\Phi}(\boldsymbol\chi,t),t\}}
{\partial (\boldsymbol\chi,t)} 
= 
\left[
\begin{array}{cc}
\frac{\partial \boldsymbol{\Phi}}{\partial \boldsymbol\chi} &
\frac{\partial \boldsymbol{\Phi}}{\partial t} \\[0.5em]
\mathbf{0}^\textrm{T} & 1
\end{array}
\right] = 
\left[
\begin{array}{cc}
\mathbf{h} & \dot{\mathbf{w}} \\[0.5em]
\mathbf{0}^\textrm{T} & 1
\end{array}
\right]
\end{equation}
where the material and spatial displacement fields are
\begin{equation}
\mathbf{W} = \mathbf{X} - {\boldsymbol\chi}\quad\textrm{and}\quad
\mathbf{w} = \mathbf{x} - {\boldsymbol\chi}
\end{equation}
$\mathbf{H}$ and $\mathbf{h}$ are the gradients of the material and spatial maps. Given the inverse of (\ref{eq::mat_map}) as
\begin{equation}
\left({\frac{\partial \{{\boldsymbol\Xi}(\boldsymbol\chi,t),t\}}
{\partial (\boldsymbol\chi,t)} }\right)^{-1}
=
\left[
\begin{array}{cc}
\mathbf{H} & \dot{\mathbf{W}} \\
\mathbf{0}^\textrm{T} & 1
\end{array}
\right]^{-1}=\left[
\begin{array}{cc}
\mathbf{H}^{-1} & -\mathbf{H}^{-1}\dot{\mathbf{W}} \\
\mathbf{0}^\textrm{T}  & 1
\end{array}
\right]
\end{equation}
then
\begin{equation} \label{eq::varphi}
\frac{\partial \{{\boldsymbol\varphi}(\mathbf{X},t),t\}}
{\partial (\mathbf{X},t)} 
=
\left(\frac{\partial \{{\boldsymbol\Phi}(\boldsymbol\chi,t),t\}}
{\partial (\boldsymbol\chi,t)}\right) \,
\left({\frac{\partial \{{\boldsymbol\Xi}(\boldsymbol\chi,t),t\}}
{\partial (\boldsymbol\chi,t)} }\right)^{-1}
=
\left[
\begin{array}{cc}
\mathbf{h}\mathbf{H}^{-1} & \dot{\mathbf{w}}-\mathbf{h}\mathbf{H}^{-1}\dot{\mathbf{W}} \\
\mathbf{0}^\textrm{T}  & 1
\end{array}
\right]
\end{equation}
It can be seen from (\ref{eq::varphi}) that the deformation gradient is
\begin{equation}
\mathbf{F} = \frac{\partial \boldsymbol\varphi}{\partial \mathbf{X}} = \mathbf{h}\mathbf{H}^{-1}
\end{equation}
Given that the physical material cannot penetrate itself or reverse the orientation of material coordinates, we have:
\begin{equation} \label{eq:det}
\textrm{det}(\mathbf{F}) = 
\frac{\textrm{det}(\mathbf{h})}
{\textrm{det}(\mathbf{H})} > 0
\end{equation}
In addition, from Eq.~(\ref{eq::varphi}), we can define the velocity of a material point $\mathbf{X}$ and the time
derivative of the deformation gradient of a point $\mathbf{X}=\textrm{const}$ as
\begin{equation}\label{eq:phy_vel}
\dot{\mathbf{u}}=\dot{\boldsymbol\varphi}=
\dot{\mathbf{w}}-\mathbf{h}\mathbf{H}^{-1}\dot{\mathbf{W}}  = \dot{\mathbf{w}}-\mathbf{F}\dot{\mathbf{W}} 
\end{equation}
and
\begin{equation}\label{eq::gard_vel}
\dot{\mathbf{F}} =
\nabla_{\mathbf{X}} \dot{\mathbf{w}} - \mathbf{h}\mathbf{H}^{-1}\nabla_{\mathbf{X}} \dot{\mathbf{W}}
\end{equation}

\section{Crack, crack front and boundary conditions in material space}

\begin{figure}[th]
\setlength{\fboxsep}{0pt}%
\setlength{\fboxrule}{0pt}%
\begin{center}
\includegraphics[width=0.5\textwidth]{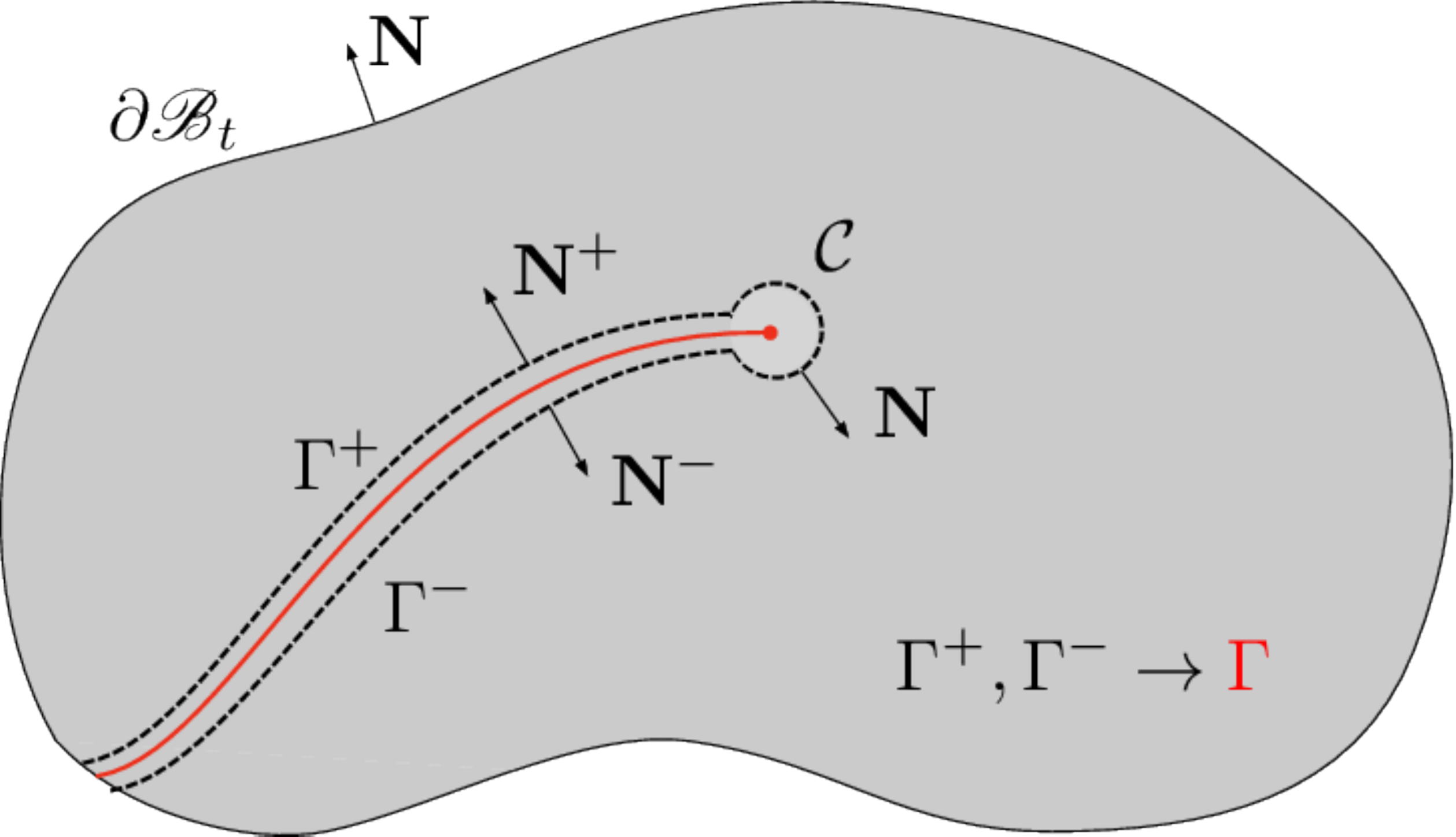}
\includegraphics[width=0.45\textwidth]{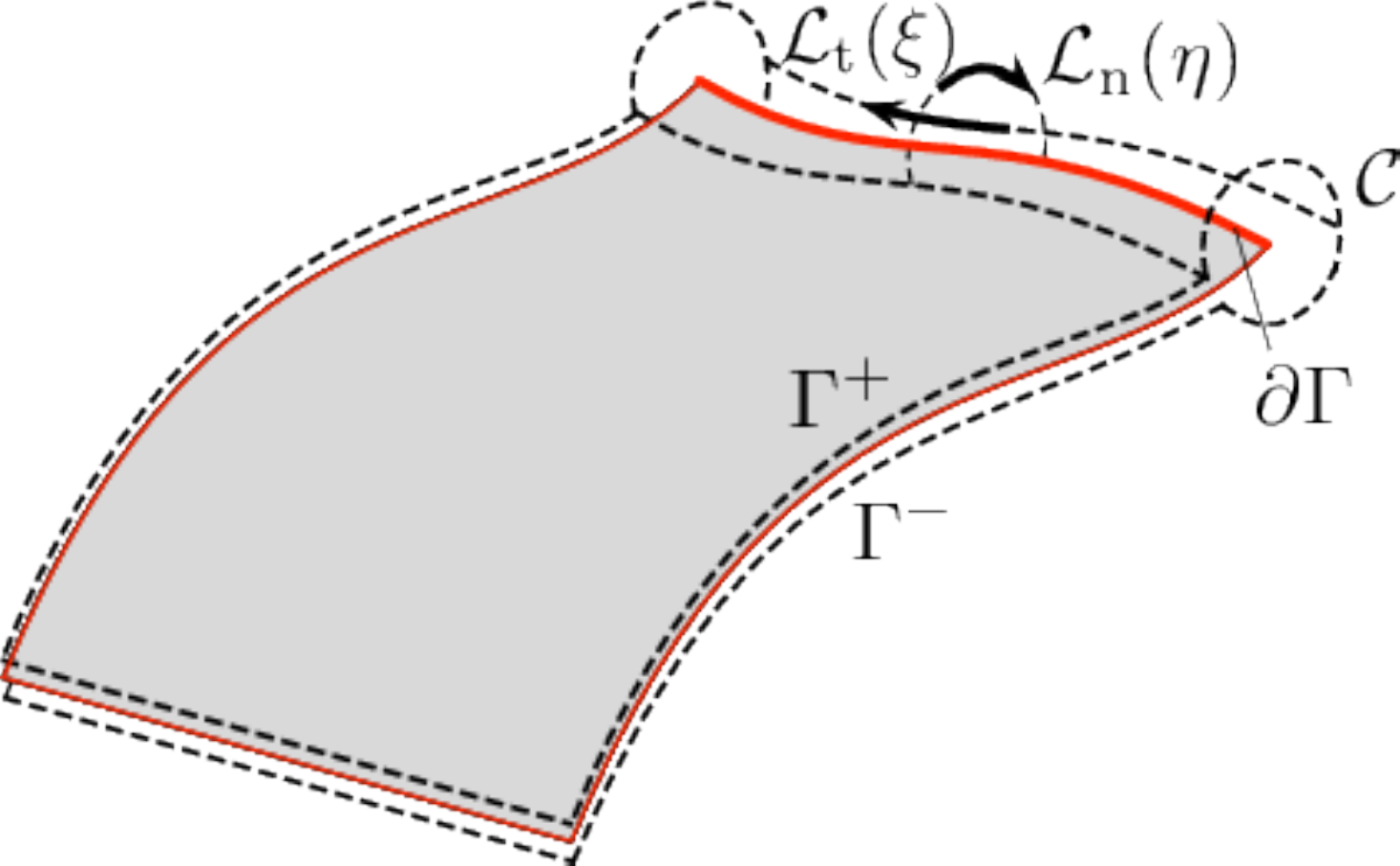}
\end{center}
\caption{Crack construction.\label{F2}}
\end{figure}
Similar to \cite{R6}, the crack surface is denoted as $\Gamma \in \mathscr{B}_t$, with a crack front $\partial\Gamma$. 
The two faces of the crack surface are given by $\Gamma^+$ and $\Gamma^-$ and $\mathcal{C}$ is a surface that encircles the crack front, see Figure~\ref{F2}.
The crack surface is obtained by the limits $\Gamma^+,\Gamma^- \to \Gamma$ and $| \mathcal{C} | \to 0$. The surface $\Gamma$ in
three dimensional space can be parameterized by coordinates $\xi$ and $\eta$, such that $\Gamma =
\Gamma(\xi,\eta)$. Defining $\mathbf{T}^{+,-}_\xi$ and $\mathbf{T}^{+,-}_\eta$ as the tangent vectors to
the surface $\Gamma^{+,-}$:
\begin{equation}
\label{eq::crack_tangent}
\mathbf{T}^{+,-}_\xi
= 
\left.
\frac{\partial \Gamma^{+,-}(\xi,\eta)}{\partial \xi}
\right|_{\eta=\textrm{const}}
\quad\textrm{and}\quad
\mathbf{T}^{+,-}_\eta=
\left.
\frac{\partial \Gamma^{+,-}(\xi,\eta)}{\partial \eta}
\right|_{\xi=\textrm{const}}
\end{equation}
the crack normal vectors $\mathbf{N}^{+,-}$ to the surfaces $\Gamma^{+,-}$ are then given as
\begin{equation}
\label{eq::crack_normal}
\mathbf{N}^{+,-}
=\mathbf{\mathbf{T}^{+,-}_\xi} \times \mathbf{T}^{+,-}_\eta=
\textrm{Spin}[\mathbf{T}^{+,-}_\xi]\mathbf{T}^{+,-}_\eta 
\end{equation}
The $\textrm{Spin}[\cdot]$ operator is introduced for convenience in calculating derivatives.

The crack surface area is given as:
\begin{equation}
A_\Gamma = 
\frac{1}{2}
(A^+_\Gamma + A^{-}_\Gamma) = 
\frac{1}{2}
\sum_{i}^{+,-}
\int_{\Gamma^i} \| \mathbf{N}^{i} \| \textrm{d}\xi \textrm{d}\eta
\end{equation}
The change of this crack surface area in the material configuration, for continuously evolving surfaces $\Gamma^+$ and $\Gamma^-$, is expressed as
\begin{equation} \label{eq::Agamma}
\begin{split}
\dot{A}_\Gamma &= 
\frac{1}{2} 
\sum^{+,-}_{i}
\int_{\Gamma^i}
\left\{
\frac{\mathbf{N}^{i}}{\| \mathbf{N}^{i} \|} 
\cdot
\left(
\textrm{Spin}\left[\mathbf{T}^{i}_\xi\right]
\frac{\partial\mathbf{T}^{i}_\eta}{\partial\mathbf{W}}
-
\textrm{Spin}\left[\mathbf{T}^{i}_\eta\right]
\frac{\partial\mathbf{T}^{i}_\xi}{\partial\mathbf{W}}
\right) 
\right\}\dot{\mathbf{W}}
\textrm{d}\xi \textrm{d}\eta\\
& = 
\frac{1}{2} 
\sum^{+,-}_{i}
\int_{\Gamma^i}
\mathbf{A}_{\Gamma^i} \cdot \dot{\mathbf{W}}
\textrm{d}S
\end{split}
\end{equation}
where, in the limit, 
$\mathbf{A}_{\Gamma^{+}}, \mathbf{A}_{\Gamma^{-}} \to \mathbf{A}_{\Gamma}$. This quantity describes
the current orientation of the crack surface. The evolution of $\mathbf{A}_{\Gamma}$ is driven by physical considerations discussed in the following sections.
It should be noted that $\mathbf{A}_{\Gamma}$ is a well defined vector for every point on the
surface $\Gamma$, including the crack front $\partial\Gamma$, and has dimensions of $m^{-1}$. 

The surface encircling the crack front $\mathcal{C}$ can be parameterised by two families
of curves, see Figure~\ref{F2}, $\mathcal{L}(\xi)_\textrm{t}=\mathcal{C}(\xi,\eta)|_{\eta=\textrm{const}}$ and 
$\mathcal{L}(\eta)_\textrm{n}=\mathcal{C}(\xi,\eta)|_{\xi=\textrm{const}}$. In the limit $|\mathcal{C}|\to0$ and
integrals over the crack front are given by
\begin{equation}
\lim_{|\mathcal{C}|\to0} 
\int_{\mathcal{C}} (\cdot) \textrm{d}S := 
\lim_{|\mathcal{C}|\to0} 
\int_{\mathcal{L}_\textrm{t}} \int_{\mathcal{L}_\textrm{n}} (\cdot) \textrm{d}S
=
\int_{\partial\Gamma} \lim_{|\mathcal{L}_\textrm{n}|\to0} \int_{\mathcal{L}_\textrm{n}} (\cdot) \textrm{d}S.
\end{equation}
The surface $\Gamma$ is created by a curve
$\mathcal{L}_\textrm{t}(\xi)$ that is swept along the path which is determined by the direction of 
$\mathbf{A}_\Gamma$.  The kinematic relationship between the
crack surface area $A_\Gamma$ and the crack front velocity $\dot{\mathbf{W}}$ on
$\mathcal{C}$ is given as
\begin{equation} \label{eq::partialAgamma}
\dot{A}_\Gamma = 
\int_{\partial\Gamma} 
\lim_{|\mathcal{L}_\textrm{n}|\to0}
\int_{\mathcal{L}_\textrm{n}} 
\mathbf{A}_\Gamma \cdot \dot{\mathbf{W}} \textrm{d}S 
=
\int_{\partial\Gamma}
\mathbf{A}_{\partial\Gamma} \cdot \dot{\mathbf{W}} \textrm{d}L.
\end{equation} 
where $\mathbf{A}_{\partial\Gamma}$ is a kinematic state variable that defines the direction of crack propagation. 
%The above equation (\ref{eq::partialAgamma}) reflect kinematic
%mechanism how the new surface area is created, i.e. by a sweep of the crack of
%crack front $\partial\Gamma$ in the material space. The crack direction
%$\mathbf{A}_{\partial\Gamma}$ is determined by physics of the problem which is
%described in the following section.
Note that the spatial velocity field $\dot{\mathbf{w}}$ is restricted in the usual manner by
the Dirichlet boundary conditions and interpenetration of surfaces $\Gamma^+$ and
$\Gamma^-$ is not admissible. 

\section{Dissipation of energy at crack front}

We consider a thermodynamic system composed of a solid elastic body with a crack.
Forces applied on the boundary of the
elastic body $\partial\mathscr{B}_t$ can do work on the system and all energy within the volume of the body can be
used to do mechanical work. Exchange of energy between the crack surface and
the exterior is restricted to the crack front. No free energy is stored at the crack front
to do mechanical work on the elastic body. Making use of Eq.~(\ref{eq:phy_vel}), the power of
external work on the elastic body is given as: 
\begin{equation}
\mathscr{P} := \int_{\partial\mathscr{B}_t} \dot{\mathbf{u}} \cdot\mathbf{t} \textrm{d}S 
= \int_{\partial\mathscr{B}_t} \left\{\dot{\mathbf{w}}\cdot\mathbf{t}-\dot{\mathbf{W}}\cdot\mathbf{F}^\textrm{T}\mathbf{t}  \right\} dS
\end{equation} 
where $\mathbf{t}$ is the external traction vector. The rate of change of internal energy of the system can be decomposed as follows:
\begin{equation} \label{eq::it_work} \dot{\mathscr{U}} :=
\dot{\mathscr{U}}_{\Gamma} + \dot{\mathscr{U}}_{\mathscr{B}_t} 
\end{equation}
where $\mathscr{U}_{\Gamma}$ is the internal crack energy and $\mathscr{U}_{\mathscr{B}_t}$ is the internal body energy. The former is defined as:
\begin{equation} 
\mathscr{U}_{\Gamma} :=
%\frac{1}{2} \sum_{i}^{+,-} \gamma A_{\Gamma^i} 
\gamma A_{\Gamma} 
\end{equation} 
where $\gamma$ is the
surface energy and has dimension $[Nm^{-1}]$. Since the new crack surface is
created by the sweep of the curve $\mathcal{L}_\textrm{t}$ along the path defined by 
$\mathbf{A}_{\partial\Gamma}$, the change of the crack surface internal
energy (noting Eq.~(\ref{eq::partialAgamma})) is expressed as:
\begin{equation} 
\dot{\mathscr{U}}_{\Gamma} :=
\frac{\textrm{d}}{\textrm{d}t}\mathscr{U}_{\Gamma_t} = \gamma \dot{A}_\Gamma=
%\frac{1}{2} \sum_{i}^{+,-} \gamma \dot{A}_\Gamma^i =
%\int_{\partial\Gamma}  \gamma \mathbf{A}_{\partial\Gamma} \cdot \dot{\mathbf{W}} \textrm{d}L
\gamma \int_{\partial\Gamma}  \mathbf{A}_{\partial\Gamma} \cdot \dot{\mathbf{W}} \textrm{d}L
\end{equation} 
%The above establishes a link between the material displacements
%$\dot{\mathbf{W}}$ and the change in crack area $\dot{A}_\Gamma$. 
Since we assume that
there is no dissipation of energy within the volume of the body, the change of
internal body energy can be expressed by 
\begin{equation}
\label{int_en}
\dot{\mathscr{U}}_{\mathscr{B}_t} :=
\frac{\textrm{d}}{\textrm{d}t}\int_{\mathscr{B}_t} \hat{\Psi}(\mathbf{F})
\textrm{d}V, 
\end{equation} 
where $\hat{\Psi}$ is the specific free energy function. Given the relation in Eq.~(\ref{eq::gard_vel}) and that 
$\textrm{d}\dot{V} = \nabla_\mathbf{X} \cdot \dot{\mathbf{W}} \textrm{d}V$, Eq.~(\ref{int_en}) can be expressed as
\begin{equation} 
\dot{\mathscr{U}}_{\mathscr{B}_t} = 
\int_{\mathscr{B}_t} \{
\mathbf{P}:\nabla_{\mathbf{X}}\dot{\mathbf{w}} +
{\boldsymbol\Sigma}:\nabla_{\mathbf{X}}\dot{\mathbf{W}} \} \textrm{d}V 
\end{equation} 
where \begin{equation} 
\mathbf{P}:=\frac{\partial \Psi(\mathbf{F})}{\partial \mathbf{F}},\quad
{\boldsymbol\Sigma} :=
\Psi(\mathbf{F})\mathbf{1}-\mathbf{F}^\textrm{T}\mathbf{P}
\end{equation} 
are the first Piola-Kirchhoff stress and Eshelby stress tensors, respectively. 
Therefore, the first law of thermodynamics, $\mathscr{P}=\dot{\mathscr{U}}_{\Gamma} + \dot{\mathscr{U}}_{\mathscr{B}_t} $,  
can be expressed as
%which satisfy the
%polyconvexity requirement, therefore ensuring the existence of minimisers in
%variational problems appearing in the framework of the finite element method.
%Making use of Eqs.~(\ref{eq:phy_vel},\ref{eq::gard_vel}) and
%$\textrm{d}\dot{V} = \nabla_\mathbf{X} \cdot \dot{\mathbf{W}} \textrm{d}V$ in
%Eq.~(\ref{eq::it_work}), the first law is expressed by
\begin{equation} \label{eq::first_law} 
\int_{\partial\mathscr{B}_t} \left\{
\dot{\mathbf{w}}\cdot\mathbf{t}-\dot{\mathbf{W}}\cdot\mathbf{F}^\textrm{T}\mathbf{t}
\right\} \textrm{d}S 
=
\int_{\partial\Gamma} \gamma
\mathbf{A}_{\partial\Gamma} \cdot \dot{\mathbf{W}} \textrm{d}L + \int_{\mathscr{B}_t} \{
\mathbf{P}:\nabla_{\mathbf{X}}\dot{\mathbf{w}} +
{\boldsymbol\Sigma}:\nabla_{\mathbf{X}}\dot{\mathbf{W}} \} \textrm{d}V \end{equation} 

In order to get a
local form of the first law, the Gauss divergence theorem is applied to the last
integral in Eq.~(\ref{eq::first_law}) resulting in the following expression 
\begin{equation} \label{eq::div_U} 
\begin{split} 
\int_{\partial\Gamma} \gamma \mathbf{A}_{\partial\Gamma}
\cdot \dot{\mathbf{W}} \textrm{d}L &= \int_{\mathscr{B}_t}
\dot{\mathbf{w}}\cdot\{ \nabla_\mathbf{X} \cdot \mathbf{P} \} \textrm{d}V +
\int_{\mathscr{B}_t} \dot{\mathbf{W}}\cdot\{ \nabla_\mathbf{X} \cdot
{\boldsymbol\Sigma} \} \textrm{d}V \\ &+
\int_{\partial\mathscr{B}_t\cup\Gamma^+\cup\Gamma^-}
\dot{\mathbf{w}}\cdot\{\mathbf{t} - \mathbf{P}\mathbf{N} \} \textrm{d}S +
\int_{\partial\mathscr{B}_t\cup\Gamma^+\cup\Gamma^-}
\dot{\mathbf{W}}\cdot\{\mathbf{F}^\textrm{T}\mathbf{t}+\boldsymbol\Sigma\mathbf{N}\}
\textrm{d}S \\ &- \int_{\partial\Gamma} \dot{\mathbf{w}} \cdot
\lim_{|\mathcal{L}_\textrm{n}|\to0} \int_{\mathcal{L}_\textrm{n}}
\mathbf{P}\mathbf{N} \textrm{d}S + \int_{\partial\Gamma} \dot{\mathbf{W}} \cdot
\lim_{|\mathcal{L}_\textrm{n}|\to0} \int_{\mathcal{L}_\textrm{n}}
\boldsymbol\Sigma
\mathbf{N}\textrm{d}S.  
\end{split} 
\end{equation} 
The spatial and material conservation laws of linear momentum balance, for any point inside the body, are expressed as:
\begin{equation}
\nabla_\mathbf{X} \cdot \mathbf{P} =\mathbf{0},\quad\nabla_\mathbf{X} \cdot
{\boldsymbol\Sigma} =\mathbf{0}
\end{equation}
Considering admissible velocity fields and stress fields in
equilibrium with external forces, we obtain the following:
\begin{equation}
\int_{\partial\Gamma} \gamma \mathbf{A}_{\partial\Gamma} \cdot \dot{\mathbf{W}}
\textrm{d}L
-
\int_{\partial\Gamma} \dot{\mathbf{W}} \cdot 
\lim_{|\mathcal{L}_\textrm{n}|\to0} \int_{\mathcal{L}_\textrm{n}} \{
\boldsymbol\Sigma\mathbf{N} \} \textrm{d}S = 0 \end{equation}
Thus, the local form of the first law is:
\begin{equation} \label{eq::first_law_local} 
\dot{\mathbf{W}} \cdot \left( \gamma \mathbf{A}_{\partial\Gamma}
-
\mathbf{G}_{\partial\Gamma} \right) 
= 0
\end{equation} 
where the configurational (or material) force at the crack front, that is driving the crack extension, is defined as:
\begin{equation}
\mathbf{G}_{\partial\Gamma}=\lim_{|\mathcal{L}_\textrm{n}|\to0} \int_{\mathcal{L}_\textrm{n}}
\boldsymbol\Sigma\mathbf{N} \textrm{d}L
\end{equation}
Eq.~(\ref{eq::first_law_local} ) represents the equilibrium condition for the crack front. One solution is trivial, i.e. that there is no crack growth ($\dot{\mathbf{W}}=\mathbf{0}$); another solution is that $\left(\gamma \mathbf{A}_{\partial\Gamma}-\mathbf{G}_{\partial\Gamma}\right)=\mathbf{0}$; a third solution is that $\dot{\mathbf{W}}$ is orthogonal to $\left(\gamma \mathbf{A}_{\partial\Gamma}-\mathbf{G}_{\partial\Gamma}\right)$. Note that the first law only defines if the crack is in equilibrium but not how it will evolve.

%and 
%there are clearly two possible solution; either that there is no crack growth ($\dot{\mathbf{W}}=\mathbf{0}$) or that the
%crack is propagating:
%\begin{equation}  
%\label{eq::first_law_local_2} 
%\gamma \mathbf{A}_{\partial\Gamma}
%=
%\mathbf{G}_\Gamma
%\end{equation} 

%Note that the material velocity $\dot{\mathbf{W}}$ or $\mathbf{A}_{\partial\Gamma}$ is not
%determined at equilibrium.  For that reason it is necessary to focus the
%attention on the second law, which set restrictions on physical equations.

Since the free energy of the elastic body can be transformed into work at the
crack front (or other forms of energy) but no energy is
stored on the crack surface that can be used to do mechanical work on the body,
%(this assumption is only valid for large scales, for small scales change of
%spatial surface curvature change system energy), 
the local variant of the second
law is simply given as: 
\begin{equation} \label{eq::second_law} 
\mathscr{D} := \gamma
\dot{\mathbf{W}} \cdot \mathbf{A}_{\partial\Gamma} \geq 0
\end{equation}
where $\mathscr{D}$ is the dissipation of energy per unit length of the crack front and is equivalent to the local change of the crack surface internal energy. This equation expresses the constraint that a physically admissible
evolution of the crack is restricted to positive crack area growth at each point of
the crack front. Healing is not thermodynamically admissible unless some other
(bio/chemo/mechanical) process takes place. Although the second law places restrictions on the
direction of crack evolution, it does not determine how $\mathbf{A}_{\partial\Gamma}$ or
$\dot{\mathbf{W}}$ evolves and therefore has to be supplemented by
a constitutive law; this is described in the next section. With this constitutive law, configurational mechanics provides a thermodynamically consistent framework which accounts for topological changes to the material body, in the form of crack propagation, and delivers the material displacement field $\mathbf{W}$
and the work conjugate configurational forces $\mathbf{G}_{\partial\Gamma}$.

\section{Material resistance}

A straightforward criterion for crack growth, in the spirit of Griffith, is proposed:
\begin{equation} \label{eq:grif1}
\phi(\mathbf{G}_{\partial\Gamma}) = 
\mathbf{G}_{\partial\Gamma} \cdot \mathbf{A}_{\partial\Gamma} - g_c/2 \leq 0
\end{equation} 
where $g_c$ is is a material parameter specifying the critical threshold of energy release
per unit area of the crack surface $\Gamma$. 

The principle of maximum dissipation for each point of the crack front states
that, for all possible Griffith-like forces $\mathbf{G}^*_{\partial\Gamma}$
that satisfy Eq.~(\ref{eq:grif1}), for a given crack configuration, the
dissipation $\mathscr{D}$ attains its maximum for the actual configurational  
force $\mathbf{G}_{\partial\Gamma}$. Therefore, we have
\begin{equation}
(\mathbf{G}_{\partial\Gamma}-\mathbf{G}^*_{\partial\Gamma})\cdot\dot{\mathbf{W}} \geq 0
\end{equation}
Using the classical method of Lagrange multipliers, we transform this into an unconstrained minimisation problem. The Lagrangian functional is therefore defined as:
\begin{equation}
\mathscr{L}(\mathbf{G}^*_{\partial\Gamma},\dot{\kappa}) = - \mathbf{G}^*_{\partial\Gamma}\cdot\dot{\mathbf{W}} + \dot{\kappa}\phi(\mathbf{G}^*_{\partial\Gamma}).
\end{equation}
and the Kuhn-Tucker conditions become
\begin{equation}
\frac{\partial\mathscr{L}}{\partial\mathbf{G}^*_{\partial\Gamma}}=-\dot{\mathbf{W}}+\dot{\kappa}\mathbf{A}_{\partial\Gamma}=0,\quad
\dot{\kappa} \geq 0,\quad\phi(\mathbf{G}^*_{\partial\Gamma}) \leq 0,\quad\textrm{and}\quad\dot{\kappa}\phi(\mathbf{G}^*_{\partial\Gamma}) = 0.
\end{equation}
In an unloading situation, i.e. $\phi<0$, we get the trivial solution of no crack growth,
\begin{equation} \label{eq::front0}
\dot{\mathbf{W}} = \mathbf{0},
\end{equation}
where the first law (\ref{eq::first_law}) is automatically satisfied and the crack
orientation $\mathbf{A}_{\partial\Gamma}$ is determined by the current crack geometry. For loading 
of a point at the crack front, $\dot{\kappa} > 0$ (and $\phi = 0$),
\begin{equation} \label{eq::front1}
\dot{\mathbf{W}} = \dot{\kappa}\mathbf{A}_{\partial\Gamma},
\end{equation}
which constrains the crack extension $\dot{\mathbf{W}}$ to be collinear to
$\mathbf{A}_{\partial\Gamma}$. Referring back to the first law
(Eq.~(\ref{eq::first_law_local})) and the crack growth criterion
(Eq.~(\ref{eq:grif1})), we therefore have:
\begin{equation} \label{eq::front2}
\gamma \mathbf{A}_{\partial\Gamma} = \mathbf{G}_\Gamma \quad \textrm{and} \quad 2 \gamma = g_c
\end{equation}
Applying the principle of maximum dissipation, the crack propagation direction is determined by the configurational
force. A thermodynamically admissible crack propagation is only possible for a
positive local change of the crack surface area (see
Eq.~(\ref{eq::second_law}). We note that $\dot{\kappa}$ has dimension of length
and represents the crack surface kinematic state variable which we can
identify as
\begin{equation}
\dot{A}_\Gamma \equiv \int_{\partial\Gamma}\dot{\kappa}\,\textrm{d}L \geq 0
\end{equation}
This is in agreement with the restriction implied by the second law,
Eq.~(\ref{eq::second_law}).  It should be noted that, for simplicity, only an
isotropic crack resistance is considered here. However, 
%since the crack orientation in the space is
%given by $\mathbf{A}_{\partial\Gamma}$, 
the current methodology could be easily extended for
anisotropic materials. 

For a general loading case, the formulation presented could result in unstable crack propagation. 
%For example, if the crack front orientation
%$\mathbf{A}_{\partial\Gamma}$ is not aligned with the configurational  force, quasi-static
%infinitesimal change of external force can result in uncontrollable crack
%propagation, i.e. finite crack area increase is needed in order to achieve
%equilibrium. 
%
Therefore, in this work, the external force is controlled such that the crack surface
evolves quasi-statically. 
%At least except some countable number of
%points on the load-displacement path. Note that for other loading scenarios
%this theory not describe what happens with crack fornt for non-equilibrated
%states, i.e. unstable crack propagation.

\section{Spatial and material discretization}

The finite element approximation is applied for both the material and physical space
\begin{equation}
\begin{array}{c}
\mathbf{X}^\textrm{h} = {\boldsymbol\Phi}(\boldsymbol\chi)\widetilde{\mathbf{X}},\quad
\mathbf{x}^\textrm{h} = {\boldsymbol\Phi}(\boldsymbol\chi)\widetilde{\mathbf{x}}\\
\mathbf{W}^\textrm{h} = {\boldsymbol\Phi}(\boldsymbol\chi)\dot{\widetilde{\mathbf{W}}},\quad
\mathbf{w}^\textrm{h} = {\boldsymbol\Phi}(\boldsymbol\chi)\dot{\widetilde{\mathbf{w}}}
\end{array}
\end{equation}
where superscript $\textrm{h}$ indicates approximation and $\left(\tilde{\cdot}\right)$ indicates nodal values. Three-dimensional domains are discretised with tetrahedral finite elements with hierarchical basis functions of arbitrary polynomial order, following the work of Ainsworth and Coyle~\cite{R12}. The higher-order approximations are only applied to displacements in the spatial configuration, whereas a linear approximation is used for displacements in the material space. The material and spatial gradients of deformation are expressed in the classical form
\begin{equation}
\mathbf{H}^\textrm{h} = \mathbf{B}_\mathbf{X}(\boldsymbol\chi)\widetilde{\mathbf{X}},\quad
\mathbf{h}^\textrm{h} = \mathbf{B}_\mathbf{x}(\boldsymbol\chi)\widetilde{\mathbf{x}}
\end{equation}
and the physical deformation gradient can be expressed as:
\begin{equation}
\mathbf{F^\textrm{h}} = \mathbf{h}^\textrm{h}(\mathbf{H}^\textrm{h})^{-1} = 
\mathbf{B}_\mathbf{x}(\mathbf{X})\widetilde{\mathbf{x}}
\end{equation}

\subsection{Crack direction}

The normal to the discretized crack surface $\Gamma^\textrm{h}$, applying the FE approximation, is given by (c.f. Eq~(\ref{eq::crack_normal})):
\begin{equation}
\label{eq::crack_normal_approx}
\mathbf{N}^\textrm{h} =
\textrm{Spin}\left[
\frac{\partial\mathbf{X}^\textrm{h}}{\partial \xi}
\right]
\frac{\partial\mathbf{X}^\textrm{h}}{\partial \eta} 
\end{equation}
where $\xi$ and $\eta$ are local coordinates of an element's triangular face on the crack surface.
This normal is constant for a linear element and is easily calculated at 
Gauss integration points for
higher-order approximations. Utilising Eq.~(\ref{eq::crack_normal_approx}) and with reference to Eq.~(\ref{eq::Agamma}), the approximation to the change in crack area can be expressed as
\begin{equation}
\begin{split}
\dot{A}_\Gamma^\textrm{h} &=\frac{1}{2}
\left\{
\A_\textrm{TRI} \int_{\textrm{TRI}}
\frac{\mathbf{N}}{\| \mathbf{N} \|} 
\cdot
\left(
\textrm{Spin}\left[\frac{\partial \mathbf{X}^\textrm{h}}{\partial \xi}\right]
\frac{\partial {\boldsymbol\Phi}}{\partial \mathbf{\eta}}
-
\textrm{Spin}\left[\frac{\partial \mathbf{X}^\textrm{h}}{\partial \mathbf{\eta}}\right]
\frac{\partial {\boldsymbol\Phi}}{\partial \xi} 
\right) 
\textrm{d}\xi \textrm{d}\eta
\right\}
\dot{\widetilde{\mathbf{W}}} \\
&= \frac{1}{2}
\left\{
\A_\textrm{TRI} \int_{\textrm{TRI}}
\mathbf{A}^\textrm{h}_\Gamma
\textrm{d}\xi \textrm{d}\eta
\right\}
\dot{\widetilde{\mathbf{W}}} =
\widetilde{\mathbf{A}}^\textrm{h}_\Gamma 
\dot{\widetilde{\mathbf{W}}}
\end{split}
\end{equation}
where $\A_\textrm{TRI}$ indicates the standard FE assembly for
the triangular surfaces of elements. The discrete matrix
$\widetilde{\mathbf{A}}^\textrm{h}_\Gamma$ has dimensions of length.

It is important to reiterate that nodes on the discretised crack surface $\Gamma^\textrm{h}$ are allowed to move in the
material configuration, but the overall body shape and its volume are constant in time. Furthermore, admissible changes in material nodal coordinates cannot change the shape of the crack faces (nor the total area of the crack surface), except for nodes on the crack front $\partial\Gamma^\textrm{h}$. 
 As a consequence, the matrix $\widetilde{\mathbf{A}}^\textrm{h}_\Gamma$ has
a number of nonzero rows equal to the number of active crack front nodes. The columns represents the contribution of the unit
nodal material velocities ($\dot{\widetilde{\mathbf{W}}}$) to changes of the crack surface area $A^\textrm{h}_\Gamma$. 

Each row of matrix $\widetilde{\mathbf{A}}^\textrm{h}_\Gamma$ can be interpreted as containing the components of
a node's material velocity which lead to a change of crack area. We
can define the vector of nodal Griffith-like forces, which represent material resistance, as
\begin{equation}
\label{eq::mat_res}
\mathbf{f}^\textrm{h}_\textrm{crack,material} := 
\frac{1}{2}\left(
\widetilde{\mathbf{A}}^\textrm{h}_\Gamma 
\right)^\textrm{T}
\mathbf{g}_c
\end{equation}
where $\mathbf{g}_c$ is a vector of size equal to the number of nodes, with $g_c$ for each component. These nodal Griffith-like forces
have dimensions of Newtons $[N]$, which are work conjugate to the material displacement $\mathbf{W}$ on the crack front.

\subsection{Residuals in spatial and material space}

The residual force vector in the discretised spatial configuration is expressed as:
\begin{equation}
\label{eq::mat_residual}
\mathbf{r}^\textrm{h}_\textrm{space} := 
\lambda\mathbf{f}^\textrm{h}_\textrm{ext,space}
-\mathbf{f}^\textrm{h}_\textrm{int,space} 
%= \mathbf{0}
\end{equation}
where $\lambda$ is an the unknown load factor and $\mathbf{f}^\textrm{h}_\textrm{ext,space}$ and $\mathbf{f}^\textrm{h}_\textrm{int,space}$ are the vectors of external and internal forces, respectively, and defined as follows
\begin{equation}
\mathbf{f}^\textrm{h}_\textrm{ext,space} := 
\A_\textrm{TRI} \int_{\textrm{TRI}} {\boldsymbol\Phi}^\textrm{T}\mathbf{t}^\textrm{h}_\lambda \textrm{d}S,\quad
\mathbf{f}^\textrm{h}_\textrm{int,space} = 
\A_\textrm{TET} \int_{\textrm{TET}} \left(\mathbf{B}_\mathbf{x}\right)^\textrm{T} \mathbf{P}^\textrm{h} \textrm{d}S
\end{equation}
where  $\A_\textrm{TET}$ indicates the standard FE assembly for tetrahedral elements.  Similarly, the residual vector
for nodal degrees of freedom in the material configuration is given by
\begin{equation}
\label{eq::material_residual}
\mathbf{r}^\textrm{h}_\textrm{material} := 
\mathbf{f}^\textrm{h}_\textrm{crack,material}
-\mathbf{f}^\textrm{h}_\textrm{int,material}
% +\tilde{\mathbf{f}}^\textrm{h}_\textrm{quality} 
%= \mathbf{0}
\end{equation}
where
$\mathbf{f}^\textrm{h}_\textrm{int,material}$ is the nodal values of configurational forces 
\begin{equation} \label{eq::nodal_resistance}
\mathbf{f}^\textrm{h}_\textrm{int,material} := 
\A_\textrm{TET} \int_{\textrm{TET}} \left(\mathbf{B}_\mathbf{x}\right)^\textrm{T} {\boldsymbol\Sigma}^\textrm{h} \textrm{d}S
\end{equation}
%and $\tilde{\mathbf{f}}^\textrm{h}_\textrm{quality}$ is a stabilisation term that has been added to 
%control the quality of the finite elements and will be discussed in detail in the
%following section. 

We note that for a continuous elastic body comprising an homogeneous material, the configurational forces within
the volume should be zero. For a discretised homogeneous body, any non-zero values are an indication of a solution error. In~\cite{R13}, this is
utilised to improve solution quality by displacing the nodal positions in the material space.
However, this improvement methodology is numerically expensive since it introduces large
non-linearities into the system which are difficult to tackle in three-dimensional
problems and can lead to convergence problems. For this reason, the nodal configurational force vector is calculated only for nodes on the crack front and an alternative approach is adopted to reduce any numerical errors in the vicinity of the crack front. An hp-adaptivity strategy is adopted utilising the hierarchical approximation basis~\cite{R12} described above and a simple mesh refinement technique, using edge-based subdivision, in the vicinity of the crack front~\cite{R14}.

Calculating the nodal configurational forces only for the nodes at the crack front, we
can write
\begin{equation}
(\widetilde{\mathbf{G}}^\textrm{h}_\Gamma)_I := (\mathbf{f}^\textrm{h}_\textrm{int,material})_I
\quad\textrm{for}\quad I \in \{I: \mathcal{N}_I\,\textrm{is crack front node}\}
\end{equation}
Finally, utilising Eq.~(\ref{eq::mat_res}), the material residual Eq.~(\ref{eq::mat_residual}) is now expressed as
\begin{equation}
\mathbf{r}^\textrm{h}_\textrm{material} = \frac{1}{2}
\left(\widetilde{\mathbf{A}}_\Gamma^\textrm{h}\right)^\textrm{T}\mathbf{g_c} - 
\widetilde{\mathbf{G}}^\textrm{h}_\Gamma 
%+ \tilde{\mathbf{f}}^\textrm{h}_\textrm{quality}
\end{equation}

\subsection{Discrete crack propagation criterion}

The crack propagation criterion for a discretised problem is expressed by a scalar product of the material residual and the nodal 
$\widetilde{\mathbf{A}}_\Gamma^\textrm{h}$, resulting in
\begin{equation} \label{eq::mat_force_error}
\widetilde{\mathbf{\phi}}^\textrm{h} := 
\widetilde{\mathbf{A}}_\Gamma^\textrm{h}\mathbf{r}^\textrm{h}_\textrm{material} = \frac{1}{2}
\widetilde{\mathbf{A}}_\Gamma^\textrm{h}(\widetilde{\mathbf{A}}_\Gamma^\textrm{h})^\textrm{T}\mathbf{g_c}
-\widetilde{\mathbf{A}}_\Gamma^\textrm{h}\widetilde{\mathbf{G}}^\textrm{h}_\Gamma < \mathbf{0},  
\end{equation} 
which can be expressed for each crack front node $I$ as:
\begin{equation} \label{eq::rac_criterion2}
g_c - 2 
\left[
\left((\widetilde{\mathbf{A}}^\textrm{h}_\Gamma)^\textrm{T}\widetilde{\mathbf{A}}^\textrm{h}_\Gamma\right)^{-1}
\widetilde{\mathbf{A}}^\textrm{h}_\Gamma\widetilde{\mathbf{G}}^\textrm{h}_\Gamma
\right]_I < 0
\end{equation}
For nodes on the crack front for which the fracture criterion is satisfied, nodes are doubled and selected faces
split. Faces are chosen based on the direction of the material force
$(\widetilde{\mathbf{G}}^\textrm{h}_\Gamma)_I$. A procedure for face
splitting is described in the following section. 

Eq~(\ref{eq::rac_criterion2}) involves a matrix inversion, where the size of the inverted matrix is equal to the
number of nodes on the crack front. A more straightforward alternative is:
\begin{equation} \label{eq::rac_criterion}
g_c - 2 \frac{(\widetilde{\mathbf{A}}^\textrm{h}_\Gamma)_I(\widetilde{\mathbf{G}}^\textrm{h}_\Gamma)_I}
{\|(\widetilde{\mathbf{A}}^\textrm{h}_\Gamma)_I\|^2}
< 0
\end{equation}
This has been tested and no significant difference to Eq~(\ref{eq::rac_criterion2}) was found. Therefore, Eq.~(\ref{eq::rac_criterion}) is used for all analyses, although a more detailed investigation is needed in the future. 

The criteria (\ref{eq::rac_criterion2}) and (\ref{eq::rac_criterion}) are a
generalization of that presented by G\"urses and Miehe in \cite{R6}, where, based
on dimensional consistency, the magnitude of the configurational nodal force is divided by
the average length of the edges adjacent to the node being split:
\begin{equation}
g_c - \frac{\|\mathbf{G}_\Gamma\|}{L_\textrm{ave}} < 0.
\end{equation}
This latter fracture criterion is equivalent to the one used here (Eq.~(\ref{eq::rac_criterion})) only for nodes with two collinear edges.

\subsection{Crack Propagation and Arc-length Method}
To trace the quasi-static dissipative loading path, we assume that the crack front is in a state of quasi-equilibrium and satisfies the crack propagation criterion Eq.~(\ref{eq::rac_criterion}), within some tolerance:
\begin{equation}
|g_c - 2
\frac{(\widetilde{\mathbf{A}}^\textrm{h}_\Gamma)_I(\widetilde{\mathbf{G}}^\textrm{h}_\Gamma)_I}
{\|(\widetilde{\mathbf{A}}^\textrm{h}_\Gamma)_I\|^2}| < \epsilon
\end{equation}
where $\epsilon$ is a parameter that depends on the characteristic mesh size. From an equilibrium configuration, the crack front is extended at each node on the crack front by one element length in the direction of the configurational force $\tilde{\mathbf{G}}_\Gamma^\textrm{h}$. This crack extension is resolved by the FE mesh using an r-adaptivity approach, whereby, for each node on the crack front, associated element faces are realigned with the predicted crack direction and a discontinuity is introduced by splitting the mesh at these faces. This process is repeated for all nodes on the crack front. The methodology for choosing an appropriate face for realignment and for subsequently splitting the mesh differs from that described in \cite{R6} and is discussed in Section~\ref{sec::realignment}. Inevitably, realignment of element faces will reduce the quality of the elements in the neighbourhood of the crack front, and this is also discussed in Section~\ref{sec::realignment}. 

For the new configuration, we adapt the external load factor $\lambda$ to achieve global equilibrium using an arc-length technique.
% This is an iterative process, whereby changes in the material and spatial displacement fields will influence the Griffith-like force and the direction of the crack extension.??????
%In order to achieve that goal, 
%for a given discretised crack front, all the nodes on the crack front 
%are in equilibrium
%if Eqns.~(\ref{eq::front1},\ref{eq::front2}) or
%Eq.~(\ref{eq::front0}) are satisfied for the loading or the unloading case,
%respectively. Based on that, 
Therefore, the system of equations for conservation of the material
and spatial momentum is supplemented by a load control equation in the form
\begin{equation} \label{eq::control}
r_\lambda = \mathbf{1} \cdot \widetilde{\mathbf{A}}^\textrm{h}_\Gamma \widetilde{\mathbf{X}}_n-
\mathbf{1} \cdot \widetilde{\mathbf{A}}^\textrm{h}_\Gamma \widetilde{\mathbf{X}}^{i+1}_{n+1} = 0,
\end{equation}
where all nodal contributions to the crack surface area are added with the use of
vector $\mathbf{1}$.  This approach of extending the crack and subsequently
determining the corresponding load factor means that we can simplify the
analysis technique by avoiding the need for developing an algorithm to deal
with the non-smooth Kuhn-Tucker loading/unloading conditions, controlling the
load such that the quasi-static crack propagation criterion is enforced at each
step.

%Since in this paper we consider quasi-static crack propagation with
%proportional load change here, this could be explored to our advantage. In such a
%such case all active crack front nodes are in neutral equilibrium. For such a case 
%an approximate numerical solution, such nodes are in quasi-equilibrium.
%Quasi-equilibrium is understand here that $|g_c - 2
%\frac{(\widetilde{\mathbf{A}}^h_\Gamma)_I^\textrm{T}(\widetilde{\mathbf{G}}^\textrm{h}_\Gamma)_I}
%{\|(\widetilde{\mathbf{A}}^h_\Gamma)_I\|^2}| < \epsilon$, that $\epsilon$ is
%some constant depending on the characteristic mesh size.  Therefore we allow
%the crack nodes, whichin the threshold $\epsilon$, to move, in order to satisfy
%constrains (\ref{eq::front2}), both for magnitude of the force and direction.
%For the propose of this paper, we call it a implicit crack propagation, since we
%attempt to correct the direction initially determined by the nodal material force
%$(\widetilde{\mathbf{G}}^\textrm{h}_\Gamma)_I$, justifying that by the fact
%that in the numerical method crack front increments are discrete and  thus finite.
%In other words for quasi-static crack propagation need of crack rotation
%express correction of numerical error, rather than non-physical crack rotation.
%Observe that utilizing load control equation (\ref{eq::control}) is no need to
%implement strong nonlinear (non-smooth) the Kuhn-Tucker loading and
%unloading conditions. 

\subsection{Linearised system of equations}
In the proposed framework, the global equilibrium solution is obtained by the Newton-Raphson method (see Algorithm~\ref{alg::alg1}), converging when 
the norms of residuals $\mathbf{r}_\textrm{spatial}$, $\mathbf{r}_\textrm{material}$
and $r_\lambda$ are less than a given tolerance. This is solved for material and
spatial displacements, as a fully coupled problem. 
We apply a standard linearisation
procedure (see for example, \cite{R13,R17} for details) to the material $\mathbf{r}^\textrm{h}_\textrm{material}$ and spatial residuals
$\mathbf{r}^\textrm{h}_\textrm{space}$, resulting in a linear system of
equations for iteration $i$ and load step $n+1$:
\begin{equation} \label{eq::lin_system}
\left[
\begin{array}{ccc}
\partial_{\widetilde{\mathbf{x}}} \mathbf{f}^\textrm{h}_\textrm{int,space} &
\partial_{\widetilde{\mathbf{x}}} \mathbf{r}^\textrm{h}_\textrm{material} &
-\partial_{\widetilde{\mathbf{x}}} \mathbf{f}^\textrm{h}_\textrm{ext,space} \\
\partial_{\widetilde{\mathbf{X}}} \mathbf{f}^\textrm{h}_\textrm{int,space} &
\partial_{\widetilde{\mathbf{X}}} \mathbf{r}^\textrm{h}_\textrm{material} &
-\partial_{\widetilde{\mathbf{X}}} \mathbf{f}^\textrm{h}_\textrm{ext,space} \\
\mathbf{0} & \partial_{\widetilde{\mathbf{X}}} r_\lambda & 0 
\end{array}
\right]
\left\{
\begin{array}{c}
\delta\widetilde{\mathbf{x}}^{i+1} \\
\delta\widetilde{\mathbf{X}}^{i+1} \\
\delta\lambda^{i+1}
\end{array}
\right\} = 
-\left[
\begin{array}{l}
\mathbf{r}^\textrm{h}_\textrm{space} \\
\mathbf{r}^\textrm{h}_\textrm{material} \\
r_\lambda
\end{array}
\right]
\end{equation}
where $\widetilde{\mathbf{X}}^{i+1} = \widetilde{\mathbf{X}}^i+\delta\widetilde{\mathbf{X}}^{i+1}$, $\widetilde{\mathbf{x}}^{i+1} = \widetilde{\mathbf{x}}^i+\delta\widetilde{\mathbf{x}}^{i+1}$ and $\lambda^{i+1} = \lambda^i+\delta\lambda^{i+1}$.

%\begin{algorithm}[t] 
%\caption{Solution procedure \label{alg::alg1} }
%\KwData{Model parameters and problem geometry}
%initialise load step $n=0$\;
%\While{body not fully fractured} {
%  \For{all nodes on crack front} {
%    \If{propagation criterion (\ref{eq::rac_criterion}) is violated}{
%      extend crack front
%    }
%  }
%  initialise iteration $i = 0$\;
%  %update refine mesh in neighborhood of crack front\;
%  %update approximation order in neighborhood of crack front\;
%  update reference configuration $\mathscr{B}_0 \leftarrow \mathscr{B}_t$\;
%  calculate residuals $\mathbf{r}_\textrm{spatial}$, $\mathbf{r}_\textrm{material}$ and $r_\lambda$\;
%  \Repeat{stop when tolerances criteria are met} {
%    calculate global tangent matrices (see (\ref{eq::lin_system}))\;
%    solve linearised system of equations\;
%    \Indp $\widetilde{\mathbf{X}}_{n+1}^{i+1} \leftarrow  \widetilde{\mathbf{X}}_{n+1}^{i} + \delta\widetilde{\mathbf{X}}^{i+1}$\;
%    $\widetilde{\mathbf{x}}_{n+1}^{i+1} \leftarrow  \widetilde{\mathbf{x}}_{n+1}^{i} + \delta\widetilde{\mathbf{x}}^{i+1}$\;
%    $\lambda_{n+1}^{i+1} \leftarrow  \lambda_{n+1}^{i+1} + \delta\lambda^{i+1}$\;
%    \Indm set iteration $i \leftarrow i + 1$\;
%    calculate residuals $\mathbf{r}_\textrm{spatial}$, $\mathbf{r}_\textrm{material}$ and $r_\lambda$\;
%  }
%  save data at load step $n$\;
%  load step $n \leftarrow n + 1$\;
%}
%\end{algorithm}

\begin{algorithm}[t] 
\caption{Solution procedure \label{alg::alg1} }
\KwData{Model parameters and problem geometry}
initialise load step $n=0$\;
\While{body not fully fractured} {
      \For{all nodes on crack front} {
      \If{crack propagation criterion (\ref{eq::rac_criterion}) is violated}{
        extend crack front
      }
    }
  initialise iteration $i = 0$\;
  %update refine mesh in neighborhood of crack front\;
  %update approximation order in neighborhood of crack front\;
  update reference configuration $\mathscr{B}_0 \leftarrow \mathscr{B}_t$\;
  \While{not converged} {
    compute tangents and residuals and\newline
    solve linearised system of equations (see (\ref{eq::lin_system}))\;
    \Indp $\widetilde{\mathbf{X}}_{n+1}^{i+1} \leftarrow  \widetilde{\mathbf{X}}_{n+1}^{i} + \delta\widetilde{\mathbf{X}}^{i+1}$\;
    $\widetilde{\mathbf{x}}_{n+1}^{i+1} \leftarrow  \widetilde{\mathbf{x}}_{n+1}^{i} + \delta\widetilde{\mathbf{x}}^{i+1}$\;
    $\lambda_{n+1}^{i+1} \leftarrow  \lambda_{n+1}^{i+1} + \delta\lambda^{i+1}$\;

     \Indm set iteration $i \leftarrow i + 1$\;
%    calculate residuals $\mathbf{r}_\textrm{spatial}$, $\mathbf{r}_\textrm{material}$ and $r_\lambda$\;
  }

  save data at load step $n$\;
  load step $n \leftarrow n + 1$\;
}
\end{algorithm}

\section{Mesh quality control}
\label{sec::mesh_quality}

%For the methodology presented here, the displacements discontinuity is
%approximated directly on the mesh by splitting the crack face nodes and
%aligning element faces with a determined direction of the crack front
%propagation. Furthermore, the solution procedure shown in
%Alg.~(\ref{alg::alg1}) allows for a change of material nodal positions. In
%general, usig the element face alignment procedure or solution procedure by
%itself without any quality control, could lead to a severe distorted mesh, as
%resulting in ill conditioned matrices and thereby allowing Newton-Raphson
%method to diverge. However, the mesh quality control presented here prevents
%these undesirable algorithmic signs. By applying this methodology we are able
%to solve large problems on parallel computers using Krylov iterative solvers
%and using standard preconditioners. These linear solvers are known to be
%sensitive for poor matrix conditioning.

%As stated above, the displacement discontinuity of the propagating crack is
%approximated directly by 
%splitting the crack face nodes and
%aligning element faces with the determined direction of the crack front extension and then splitting the mesh. 
Extension of the crack front can result in poor quality elements in the vicinity of the crack front, including inverted or
severely distorted elements, that can make the problem difficult or impossible to solve. To address this, we apply a similar approach to the one
proposed by Scherer et al. \cite{R13}. The key challenge is to enforce
constraints to preserve element quality for each Newton-Raphson
iteration, without influencing the physical response. Thus, we introduce a measure of element quality for tetrahedral elements in terms of their shape and use this to drive mesh improvement. Here we restrict ourselves to movement of nodes, with no changes to element topology. By applying this methodology we are able
to solve large problems on parallel computers using Krylov iterative solvers with standard preconditioners.

The dihedral angles formed between the faces of a tetrahedron have been shown
to be one of the most influential properties in terms of solution accuracy
\cite{R25}. Large dihedral angles result in interpolation errors and small dihedral
angles affect the conditioning of the stiffness matrix. The minimum sine of the
dihedral angles has been used in \cite{R25}, however, this is a non-smooth function
that cannot be used with a Newton-Raphson solver. To overcome this, an
alternative volume-length measure is proposed. Although this measure does not
directly measure dihedral angles, it has been shown to be very effective at
eliminating poor angles, thus improving stiffness matrix conditioning and
interpolation errors \cite{R15,R16}. As the volume-length measure is a smooth
function of vertex positions and its gradient is straightforward and
computationally cheap to calculate, it is ideal for the problem at hand. The
volume-length quality measure is defined as
\begin{equation} \label{eq::quality_mm}
q(\mathbf{H}^\textrm{h}) := 6\sqrt{2} \frac{V_0}{l_{\textrm{rms},0}^3}\frac{\textrm{det}(\mathbf{H}^\textrm{h})}{dl_{\textrm{rms}}^3(\mathbf{H}^\textrm{h})} 
= q_0b(\mathbf{H}^\textrm{h})
,\quad
b(\mathbf{H}^\textrm{h}) := \frac{\textrm{det}(\mathbf{H}^\textrm{h})}{dl_\textrm{rms}^3(\mathbf{H}^\textrm{h})}
\end{equation}
where $q_0$, $V_0$ and $l_{\textrm{rms},0}$ are the element quality, element volume and root mean square of the element's edge lengths respectively, in the reference configuration. $\mathbf{H}^\textrm{h}$ is the material deformation gradient, $b$ is measure of change of element quality, relative to the reference configuration, and $dl_{\textrm{rms}}=l_\textrm{rms}/l_{\textrm{rms},0}$ is the stretch of $l_{\textrm{rms},0}$ from the reference configuration. $q_0$ is normalised so that an equilateral element has quality $1$ and a degenerate element (zero volume) has
quality $0$. Furthermore, $b=1$ corresponds to no change and $b=0$ is a change leading to a degenerate element. An element edge length in the material configuration is expressed as
\begin{equation}
l_j(\mathbf{H}^\textrm{h}) := \sqrt{
\Delta {\boldsymbol\chi}_j^\textrm{T}(\mathbf{H}^\textrm{h})^\textrm{T}\mathbf{H}^\textrm{h}\Delta {\boldsymbol\chi}_j
}
\end{equation}
where $\Delta {\boldsymbol\chi}^j$ and is the distance vector of edge $j$ in the
reference configuration. Thus $l_\textrm{rms}$ is calculated as
\begin{equation}
l_\textrm{rms} :=
\sqrt{\frac{1}{6}
\sum_{j=1}^{6} l_j^2 }
 =  l_{\textrm{rms},0}dl_\textrm{rms}
\end{equation}
To control the quality of elements, we enforce an admissible deformation $\mathbf{H}^\textrm{h}$
such that
\begin{equation} \label{eq::quali_const}
b(\mathbf{H}^\textrm{h})>\gamma \quad \textrm{for}\; \gamma \in (0,1)
\end{equation}
In practice, $0.1 < \gamma <
0.5$ gives good results. This constraint on $b$ is enforced  by applying a volume--length log--barrier function \cite{R13} defined as
\begin{equation} \label{eq::mesh_energy}
%\mathcal{B} := \sum_{e=0}^\mathcal{N} \frac{q^2}{2(1-\gamma)} - \log(q_e-\gamma),\quad
\mathcal{B} := \sum_{e=0}^\mathcal{N} \frac{b_e^2}{2(1-\gamma)} - \ln(b_e-\gamma)
\end{equation}
where $\mathcal{B}$ is the barrier function for the change in element quality in the material configuration. It can be seen that the Log--Barrier function
rapidly increases as the quality of an element reduces, and tends to infinity
when the quality approaches the barrier $\gamma$, thus achieving our aim of penalising the
worst quality elements. 

In order to build a solution scheme that incorporates a stabilising force that controls element quality, we define a pseudo `stress' at the element level as a counterpart
of the first Piola--Kirchhoff stress as follows
\begin{equation} 
\mathbf{Q} :=  
\frac{\partial \mathcal{B} }{\partial \mathbf{H}^\textrm{h} } = 
\textrm{det}(\mathbf{H}^\textrm{h})
\frac{l_{\textrm{lrms},0}^3}{l_{\textrm{lrms}}^3}
 \left(
\frac{b}{1-\gamma} + \frac{1}{b-\gamma}
\right) \widetilde{\mathbf{Q}},
\end{equation}
where matrix
$\widetilde{\mathbf{Q}}$ is defined as follows
\begin{equation}\label{eq::quali_ene}
\widetilde{\mathbf{Q}} := 
(\mathbf{H}^\textrm{h})^{-\textrm{T}}
-
\frac{1}{2}
\frac{1}{l_{\textrm{lrms}}^2}
\sum_{j=1}^6 
\Delta {\mathbf{X}}^j
(\Delta {\boldsymbol\chi}^j)^\textrm{T}.
\end{equation}
It is worth noting that $\mathbf{Q}$ should be a zero matrix for a purely volumetric change or rigid body movement of
a tetrahedral element. 

It is now possible to compute a vector of nodal pseudo `forces' associated with $\mathbf{Q}$ as
\begin{equation} \label{eq::rhs_mesh}
\mathbf{f}^\textrm{h}_\textrm{quality} =
\A_\textrm{TET} 
\int_{\textrm{TET}} 
\alpha
\mathbf{B}^\textrm{T}_\mathbf{X} \mathbf{Q}
\textrm{d}V 
\end{equation} 
where $\alpha$ is a solution parameter that can vary from $0$ to $1$ but is typically set to $1$.  Thus, the global vector of nodal material residual forces (Eq.~(\ref{eq::material_residual})) is now augmented with this additional stabilisation term $\mathbf{f}^\textrm{h}_\textrm{quality}$ to give:
\begin{equation}
\mathbf{r}^\textrm{h}_\textrm{material} := 
\mathbf{f}^\textrm{h}_\textrm{crack,material}
-\mathbf{f}^\textrm{h}_\textrm{int,material}
+\mathbf{f}^\textrm{h}_\textrm{quality} 
\end{equation}
Starting from an equilibrium state (i.e. configurational forces and griffith-like forces are equal and the arc-length control (Eq.~(\ref{eq::control})) is satisfied), a subsequent load step will generally result in a loss of equilibrium and mesh deformation will occur, with 
%If the crack front is in equilibrium (i.e. configurational forces and griffith-like forces are equal) and the arc-length control Eq.~(\ref{eq::control}) satisfied, then
%$b=1$ and $\mathbf{Q} = \mathbf{0}$. If, for a
%subsequent Newton-Raphson iteration, equilibrium is not satisfied, mesh deformation will occur but 
changes in element quality controlled by the volume--length log--barrier function. Thus, for any arbitrary perturbation from equlibrium, by $\delta
\widetilde{\mathbf{X}}$, $\delta \widetilde{\mathbf{x}}$ or $\delta\lambda$,
equilibrium would be recovered by the Newton-Raphson procedure irrespective of $\alpha$. However, if $\alpha=0$, there is no control of element quality. In future, it is possible for $\alpha$ to also reflect the magnitude of the out-of-balance material force, providing a damping effect where the current iteration is a long way from equilibrium. To date, this has not been required.

\section{Determination of critical faces and mesh realignment}
\label{sec::realignment}
\begin{figure}[th]
\setlength{\fboxsep}{0pt}%
\setlength{\fboxrule}{0pt}%
\begin{center}
\includegraphics[width=1.\textwidth]{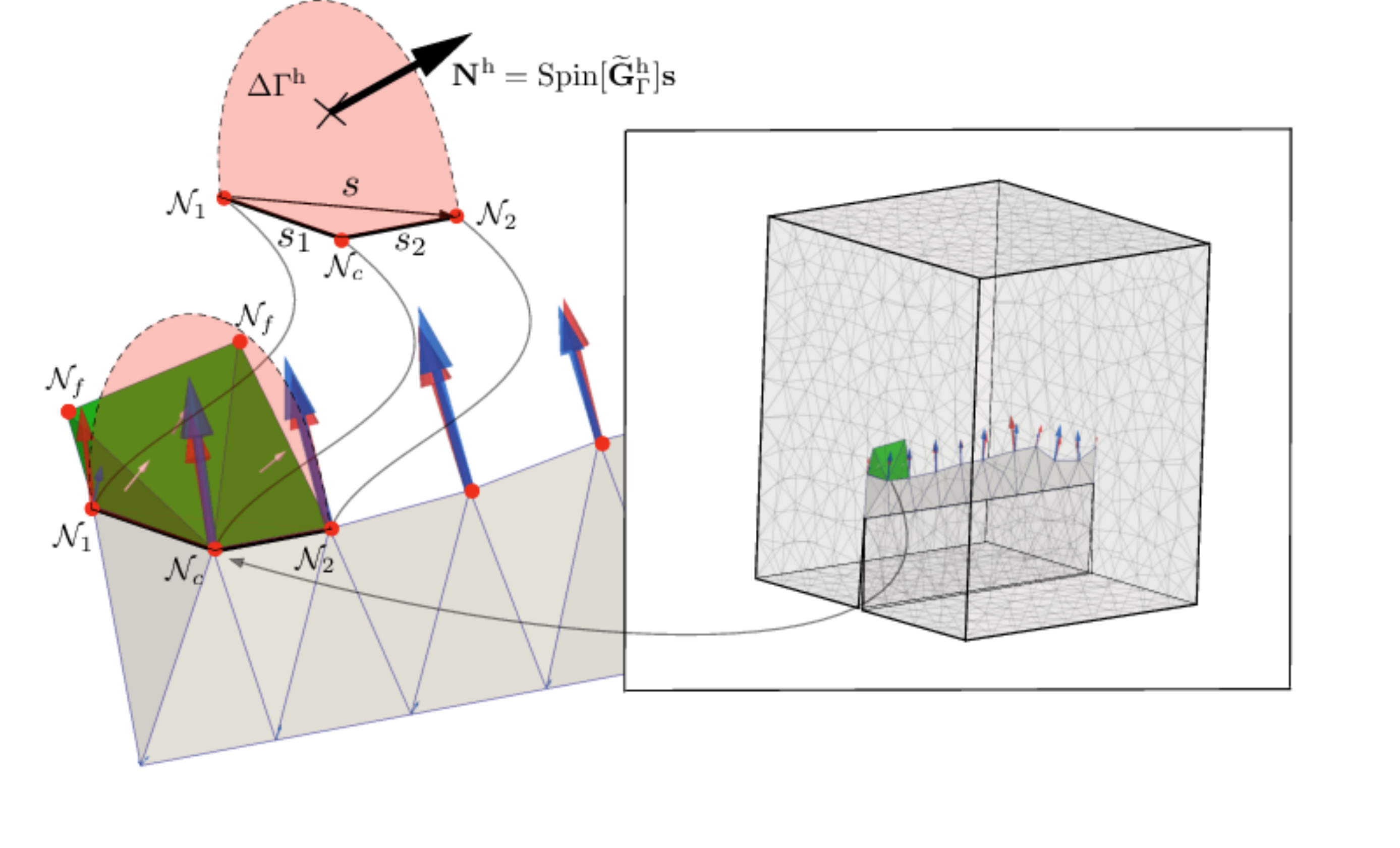}
\end{center}
\caption{Crack front extension. Green element faces represent admissible crack extension before projection on predicted crack surface extension $\Delta\Gamma^\textrm{h}$. Blue arrows represent nodal configurational forces and red arrows represent nodal Griffith-like forces. \label{F3}}
\end{figure}

To determine which element face will be aligned with the predicted crack front extension and then split in order to create the crack extension, first all nodes on the current crack front are ordered according to the degree with which they violate the discrete crack propagation criterion~(\ref{eq::rac_criterion}). Taking each of these nodes in turn, we consider a vector $\mathbf{s} = \mathbf{s}_1+\mathbf{s}_2$ on the crack front associated with node $\mathcal{N}_c$, as shown in Fig.~\ref{F3}. Next, the normal of the new crack extension, $\Delta\Gamma^\textrm{h}$, is determined as follows:
\begin{equation} \label{eq::carck_normal}
\mathbf{N}^\textrm{h} = \textrm{Spin}[\widetilde{\mathbf{G}}^\textrm{h}_\Gamma]\mathbf{s}
\end{equation}
As an example, Fig.~\ref{F3} illustrates three faces associated with node $\mathcal{N}_c$ as well as the predicted crack surface $\Delta\Gamma^\textrm{h}$ defined by $\mathbf{N}^\textrm{h}$. It is important to emphasise that, in general, there will be more than one admissible set of faces that are associated with node $\mathcal{N}_c$, but for simplicity only one set is highlighted in Fig.~\ref{F3}. Only edges and faces which are connected directly to node $\mathcal{N}_c$ are considered. Edges and faces which are part of the body's surface or the existing crack surface are excluded from these considerations as the crack can only propagate in the volume of the body.

For each admissible set of faces, `free' nodes $\mathcal{N}_f$ can be identified that can be moved so that the faces are aligned with the predicted crack extension $\Delta\Gamma^\textrm{h}$. Each admissible set of faces is tested and that set which results in the least worse element quality (Eq.~(\ref{eq::quality_mm})) is aligned with $\Delta\Gamma^\textrm{h}$, nodes are duplicated and the faces are split. The last part of the process is mesh smoothing in the vicinity of the crack front, utilising the methodology described previously.

%We consider only edges
%or faces where one of the nodes is connected to the node $\mathcal{N}_c$. From this
%consideration are excluded the cycles on the tree, i.e. self intersections, since such
%cycles projected on the new crack surface resulting in zero-volume tetrahedral
%elements. Admissible tree does not comprise edges and faces which are part of the
%surface or the crack surface, this is easy explained by fact that crack is propagating
%in the volume. 

%, see Eq.~(\ref{eq::quali_ene}).
%The problem of mesh energy minimisation is solved for the small patch of
%elements, using standard FE procedure, where  the right hand residual  is
%calculated using Eq.~\ref{eq::rhs_mesh}.

In the authors' implementation, the above methodology is achieved by constructing an undirected graph tree for all admissible crack surface extensions for each each node $\mathcal{N}_c$. The vertices of the tree represent element edges and lines of the tree represent element faces. The starting vertex is edge $s_1$ and the last vertex is edge $s_2$. In the case that node $\mathcal{N}_c$ is on the surface of the body, it is only associated with one edge, i.e. $\mathbf{s} = \mathbf{s}_1$ and the last vertex of the tree can be any edge on the body surface connected to node
$\mathcal{N}_c$. Each path through the tree represents a possible set of faces that can be realigned and split to form the crack extension. 

Once the tree is built, a search over all branches is carried out, i.e. all possible
paths which link the two adjacent crack front edges are searched for the set of faces that, 
when aligned with the predicted crack extension $\Delta\Gamma^\textrm{h}$, lead to the least worse quality element (using the volume--length quality measure (Eq.~\ref{eq::quality_mm})). Thus, this approach is optimal in minimising any possible deterioration in mesh quality as a result of mesh realignment. 
%Note that faces in selected
%branch not necessary initially create smooth crack surface, very often it can
%be quite opposite, but those one which projected preserve good quality mesh in
%the sense of Eq.~(\ref{eq::mesh_energy}).

%\begin{algorithm}[!h] 
%\caption{Projecting faces \label{alg::alg5} }
%\KwData{CrackingNode, FacesForSplitting and $\widetilde{\mathbf{G}}^\textrm{h}_\Gamma$}
%FacesForSplitting $\leftarrow$ \emph{project\_faces}(CrackingNode,$\widetilde{\mathbf{G}}^\textrm{h}_\Gamma$)\;
%\tcc{patch elements for improvement}
%Nodes $\leftarrow$ \emph{get\_adjacencies}(Faces)\;
%Elems $\leftarrow$ \emph{get\_adjacencies}(Nodes)\;
%Nodes $\leftarrow$ \emph{get\_adjacencies}(Elems)\;
%Elems $\leftarrow$ \emph{get\_adjacencies}(Nodes)\;
%\emph{improving\_elems\_in\_patch}(Elems)\;
%\end{algorithm}
Following realignment of the element faces, node $\mathcal{N}_c$ and the selected element faces are
doubled, creating two new surfaces for the crack extension. Next, mesh smoothing is undertaken for a patch of elements in the vicinity of node 
$\mathcal{N}_c$ utilising the volume-length log-barrier function described in Section~\ref{sec::mesh_quality} to ensure mesh quality. This has been found to be computationally efficient and in practice its execution takes an insignificant fraction of the total
solution time. This process was found to be very robust and works well for meshes with varying element size and/or with initially poor quality. It is also worth noting that this procedure of crack extension and mesh smoothing could be implemented in virtually any FE system as an isolated and autonomous procedure since it does not involve any enrichment nor does it change mesh
connectivity and only involves doubling of nodes and element faces. 

\section{Implementation and Numerical Examples}

\begin{figure}[th]
\setlength{\fboxsep}{0pt}%
\setlength{\fboxrule}{0pt}%
\begin{center}
\includegraphics[width=0.6\textwidth]{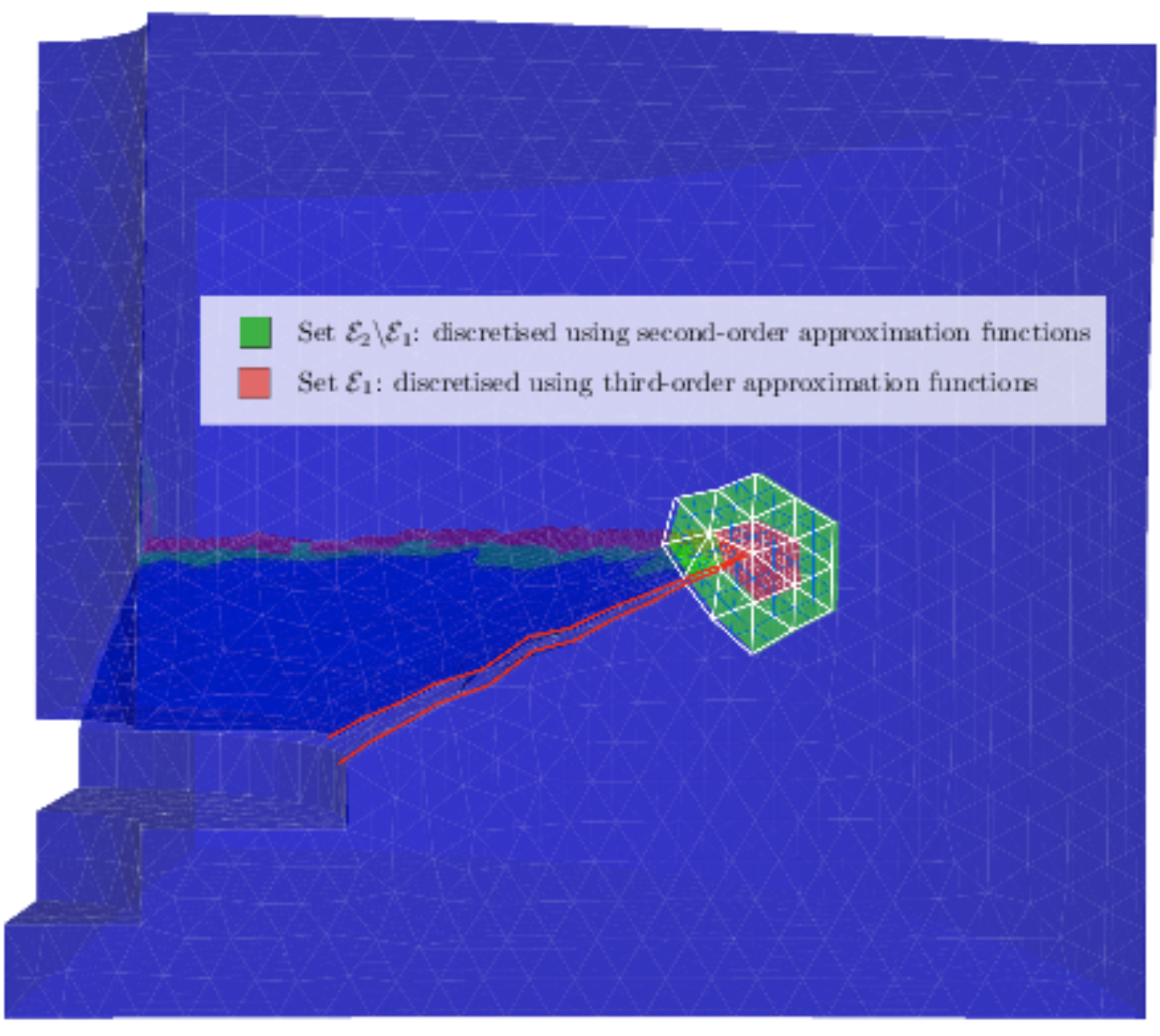}
\end{center}
\caption{hp-refinement at crack front \label{F4a}}
\end{figure}

All problems are discretised using 3D tetrahedral elements. An adaptive mesh refinement strategy is adopted that includes both h-refinement, using an edge-based splitting algorithm \cite{R14}, and p-refinement using hierarchic basis functions identified above~\cite{R12}. First, all elements $\mathcal{E}_1$ adjacent to the crack
front are selected, then a larger set of elements $\mathcal{E}_2$
adjacent to and including set $\mathcal{E}_1$ are selected. All elements in set
$\mathcal{E}_2$ are subject to edge-based splitting. The spatial nodal positions are discretised using higher-order approximations in the vicinity of the crack front. The spatial nodal positions in all elements in 
$\mathcal{E}_2 \backslash \mathcal{E}_1$ are discretised using second-order approximation functions and elements in 
$\mathcal{E}_1$ using third-order. This refinement strategy is automated so that the mesh is locally refined while the crack is propagating. This is illustrated in Fig~\ref{F4a}. As the crack front moves forward, elements are removed from set $\mathcal{E}_2$ and elements revert to their original state.

The solution strategy presented in this paper is implemented for parallel shared
memory computers, utilising open-source libraries. MOAB,  
a mesh-oriented database \cite{R18}, is used to store mesh data, including input and output
operations and information about mesh topology. PETSc (Portable, Extensible Toolkit
for Scientific Computation \cite{R19}) is used for parallel
matrix and vector operations, the solution of linear system of
equations and other algebraic operations.
ParMetis \cite{R20} is used for parallel mesh partitioning, using the PETSc
native interface.

%Element faces of the coarse mesh are aligned with the predicted crack direction. However, in order to improve the quality of the crack front approximation, only the element faces of the refined mesh are split as and when the discrete propagation
%criterion (see Eq.~(\ref{eq::rac_criterion})) is met.

\begin{figure}[th]
\setlength{\fboxsep}{0pt}%
\setlength{\fboxrule}{0pt}%
\begin{center}
\includegraphics[width=0.49\textwidth]{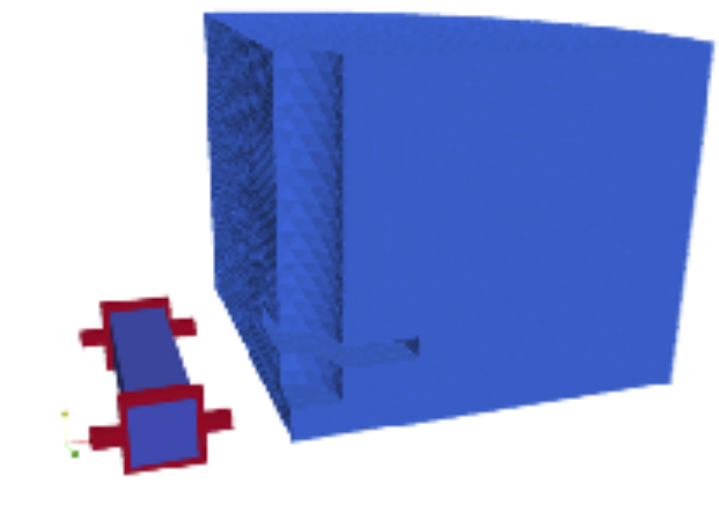}
\end{center}
\caption{Geometry and relative size comparison for torsion and pull-out test \label{F4}}
\end{figure}

To demonstrate the proposed modelling framework, two numerical examples are presented: torsion test
and pull-out test. For both examples, a quasi-brittle material is
considered, i.e. concrete. The torsion test represents the analysis of a relatively small physical sample whereas
the pull-out test represents a much larger problem, see Fig.~\ref{F4}.
These two tests not only provide a means of verifying the numerical robustness of the presented
methodology but also to illustrate the limits of the Griffith-like fracture theory adopted in this paper.

\subsection{Pull out test}

This problem considers the pull-out of a steel anchor embedded in a
concrete cylindrical block, which is surrounded by a steel ring. The problem was simulated by 
Duan \& Areias,  Belytschko \cite{R8,R9} and Gasser \& Holzapfel \cite{R10}.
All geometrical data has been obtained from \cite{R8,R9,R10}. Young's modulus $E=30000\, \textrm{N/mm}^2$, Poison ratio $\nu=0.2$ and
Griffith energy $G_f=0.106\, \textrm{N/mm}$. Following \cite{R8}, the anchor itself is not explicitly modelled; 
instead a uniform prescribed vertical displacement is applied to the concrete face corresponding to the upper face of the anchor's plate. Only one quarter of the specimen is modelled.

\begin{figure}[htb]
\setlength{\fboxsep}{0pt}%
\setlength{\fboxrule}{0pt}%
\begin{center}
\includegraphics[width=1.\textwidth]{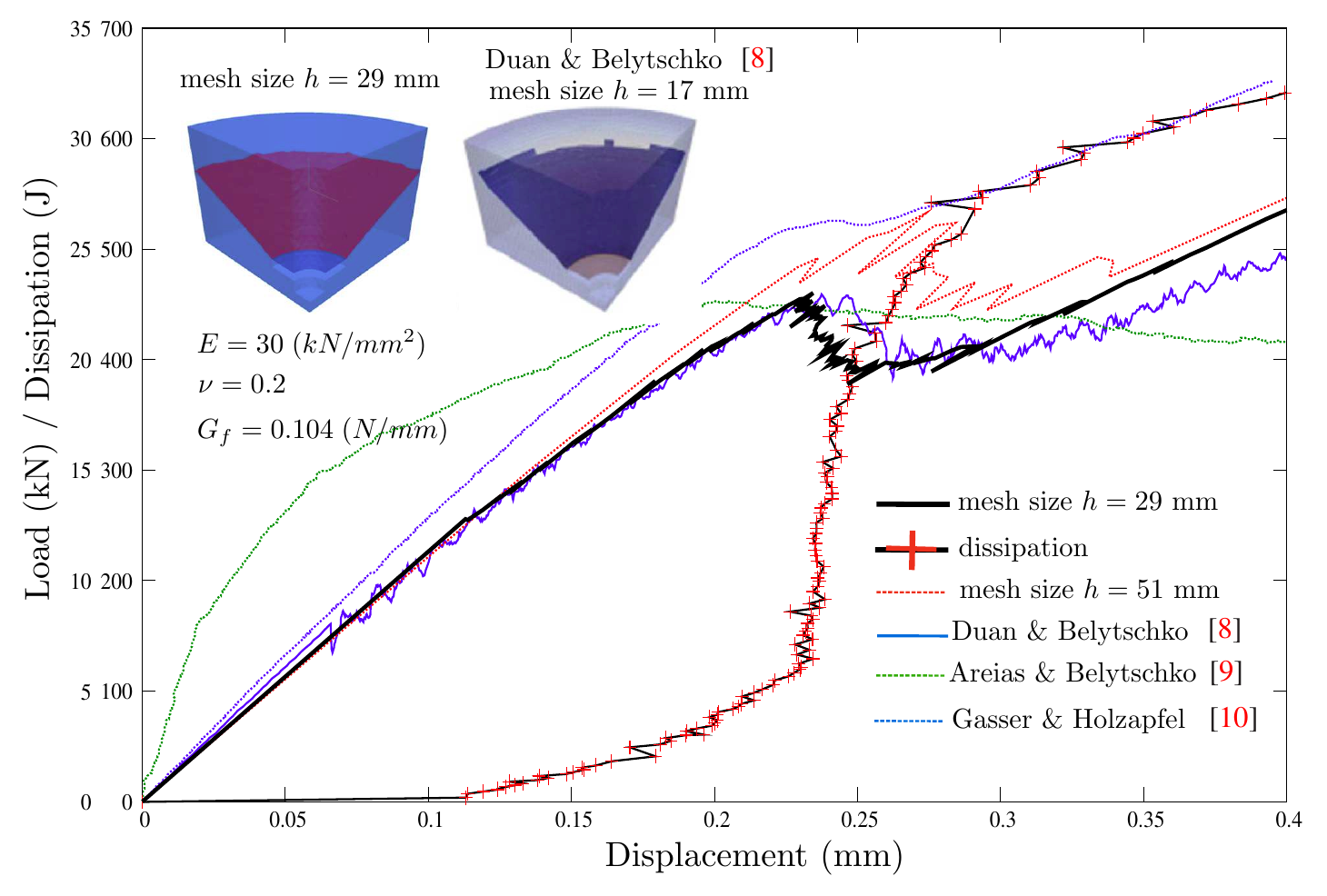}
\end{center}
\caption{Load-displacement path and energy dissipation for pull-out problem \label{F5}}
\end{figure}

A coarse mesh comprising $2995$ nodes, with a characteristic mesh size of $51$mm, and a fine mesh comprising $14075$ nodes, with a mesh size of $29$mm, have been used in the analyses. A neo-Hookean material was used initially, but both displacements and
strains were found to be small and there was no noticeable difference compared to the solution assuming a linear elastic material was observed. Therefore, all results presented here are for a linear elastic material.

\begin{figure}[htb]
\setlength{\fboxsep}{0pt}%
\setlength{\fboxrule}{0pt}%
\begin{center}
\includegraphics[width=0.49\textwidth]{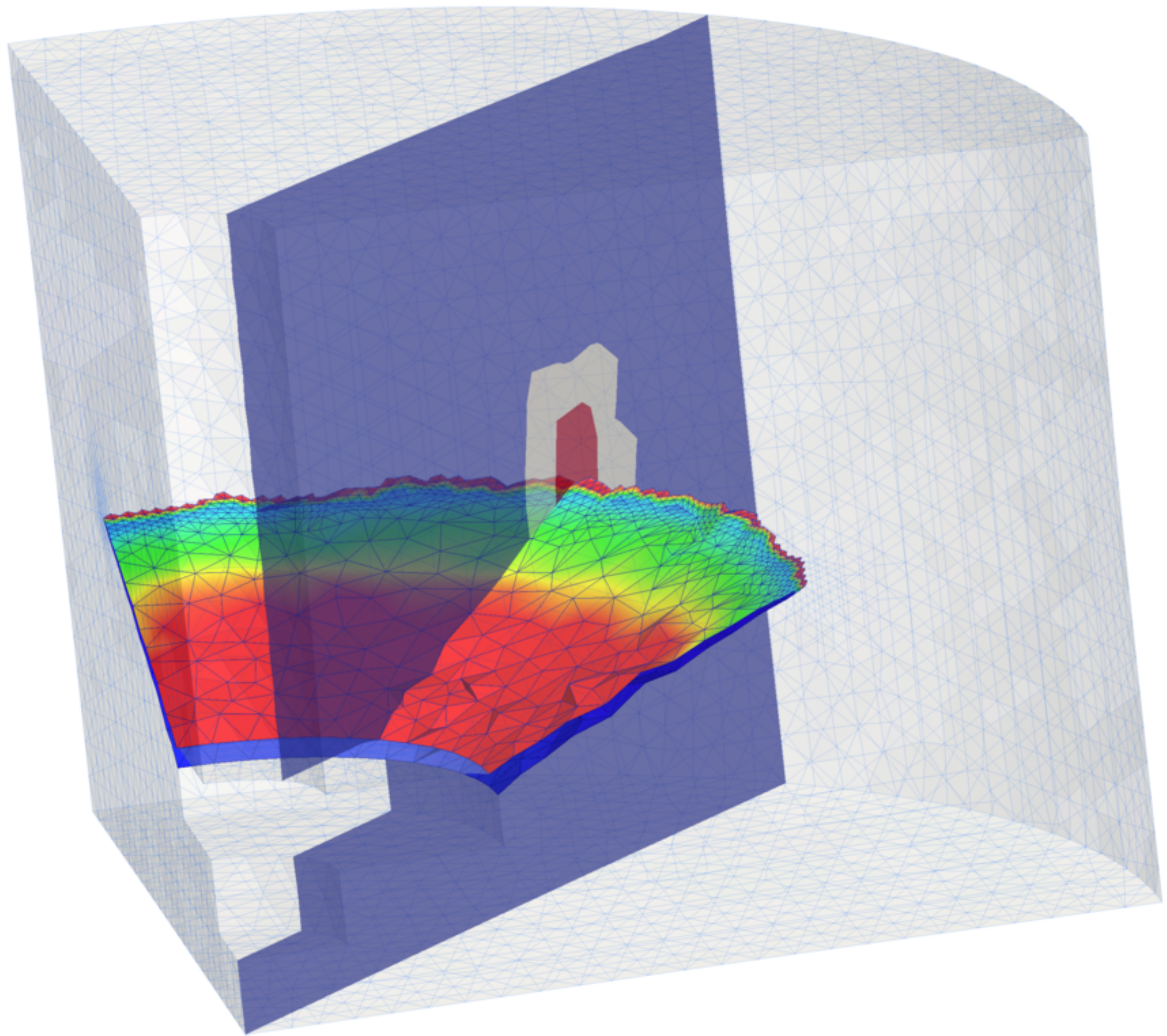}
\includegraphics[width=0.49\textwidth]{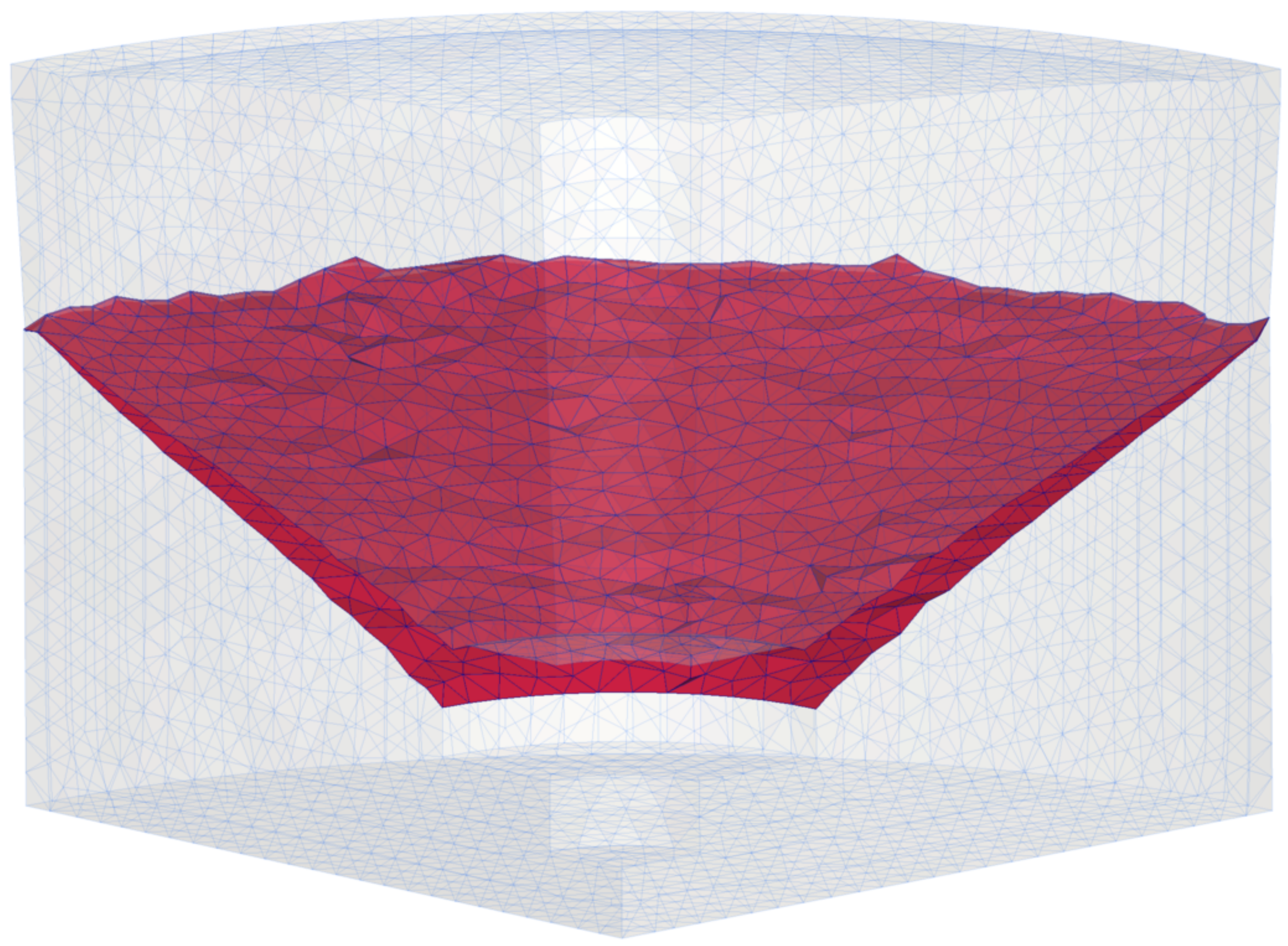} \\
(a)\qquad\qquad\qquad\qquad\qquad\qquad\qquad\qquad\qquad(b)
\end{center}
\caption{(a) Propagating crack with contour plot on crack surface of strain energy density and contour plot on vertical slice of polynomial order (blue: linear, white: quadratic, red: cubic). (b)
Final crack surface. Follow this \href{http://youtu.be/58vM_44-cOU}{\texttt{link}} for animation of the results for this problem.
\label{F6}}
\end{figure}

%In this example $\alpha_\textrm{min} = 1$ and $alpha = 0$, thus $\beta =
%\alpha_\textrm{min} = const.$ The $\beta$ parameter control material nodal
%displacements at crack front. If this value is too small it will effect in
%convergence problems, large values make impact on quality crack front
%approximation and this results in oscillations of load-displacement path. This
%could be mitigated by refining mesh at the crack front, see 
Fig.~\ref{F5} shows the load-displacement response for the two different discretisations. The
coarse mesh analysis was undertaken without any mesh refinement at the crack
front, whereas the fine mesh analysis included hp-refinement as described earlier. It can be seen that in the latter case, the oscillations in the response are smaller in magnitude. 

Fig.~\ref{F5}, also compares the load-displacement paths for simulations of the same problem reported by Gasser \& Holzapfel \cite{R10}, Areias
\& Belytschko \cite{R9} and Duan \& Belytschko \cite{R8}.
All these simulations utilise a cohesive crack methodology together with enrichment functions to approximate
displacement discontinuities.  The cohesive crack
methodology assumes that significant energy is dissipated in the fracture process zone, in the vicinity of the crack
front, associated with meso-scale cracking and incorporated into the model by a
fictitious extension of the crack front with cohesive forces. The work of these cohesive forces
is equal to the work dissipated in the fracture process zone. In cohesive models, energy is dissipated by crack opening; in contrast, in the present work, it is assumed that energy is exclusively dissipated in the process
of creating new crack area. The present approach is justified by
the assumption that the volume of the fracture process zone (where meso-cracking takes place) is
significantly smaller than the total volume of the body and so it can be assumed that
energy dissipation is restricted to the crack front (note that aggregates are approximately two orders of magnitude smaller than the diameter of the specimen). This explains the good correlation of the present results with those presented by  Duan \&
Belytschko, see Fig.~\ref{F5}.

%If the response $Y$ (e.g. reaction force on the load displacement path) of
%a geometrically similar system, is expressed as a function of a characteristic size
%$D$, the response of the structure is given by $Y=Y_0f(D)$, where $f(D)$ is
%expressed by the scaling law and $Y_0$ is solution obtained at some scale. In
%general, the solution for scaling functions has a form of $f(D) = (D/c_1)^s$, where $c_1$
%is a constant which is always implied as a unit of length measurement
%\cite{R21}. The Griffith-like fracture theory implies that $s=-\frac{1}{2}$,
%however when significant volume of fracture process zone is subjected to
%meso-cracking, because the large size of its inhomogeneities, then the above
%scaling function is no longer applicable. In such cases, a nonlinear quasi-brittle size effect
%take place. However for the large samples, e.g. the small size of concrete inhomogeneities
%the brittle fracture theory gives meaningful results, without need of artificial
%scaling of the material parameters. Note that for the pull-out test size of aggregates
%approximately two order times smaller than diameter of concrete sample. This
%explains good correlation of our result and those presented by  Duan \&
%Belytschko, see Fig.~\ref{F5}.

It is useful to observe that, in the case of Gasser \& Holzapfel \cite{R10}, the peak
strength matches the results of Duan \& Belytschko. In the case of Areias \&
Belytschko \cite{R9}, the results show an over prediction of strength compared to all other results.  It is important to recall that the present
model is based only on three material parameters, i.e. two for the elastic
solid and the third is the Griffith energy release rate; no additional crack
separation law is needed. In spite of this limited number of material parameters, good agreement with the other results is achieved.

\subsection{Torsion test}

An experimental study by Brokenshire~\cite{R23} of a torsion test of a plain concrete notched prismatic beam ($400\textrm{mm}\times 100\textrm{mm}\times 100\textrm{mm}$) is considered. The experimental procedure and full details of the boundary conditions and dimensions are described in Jefferson et al.~\cite{R22}. The notch is placed at an oblique angle across the beam and extends to half the depth. The beam is placed in a steel loading frame, supported at three corners and loaded at the fourth corner.
The experiment used aggregates with a maximum size of $10$ mm; the characteristic size
of the specimen can be considered to be $100\textrm{mm}$.

\begin{figure}[htb]
\setlength{\fboxsep}{0pt}%
\setlength{\fboxrule}{0pt}%
\begin{center}
\includegraphics[width=1.0\textwidth]{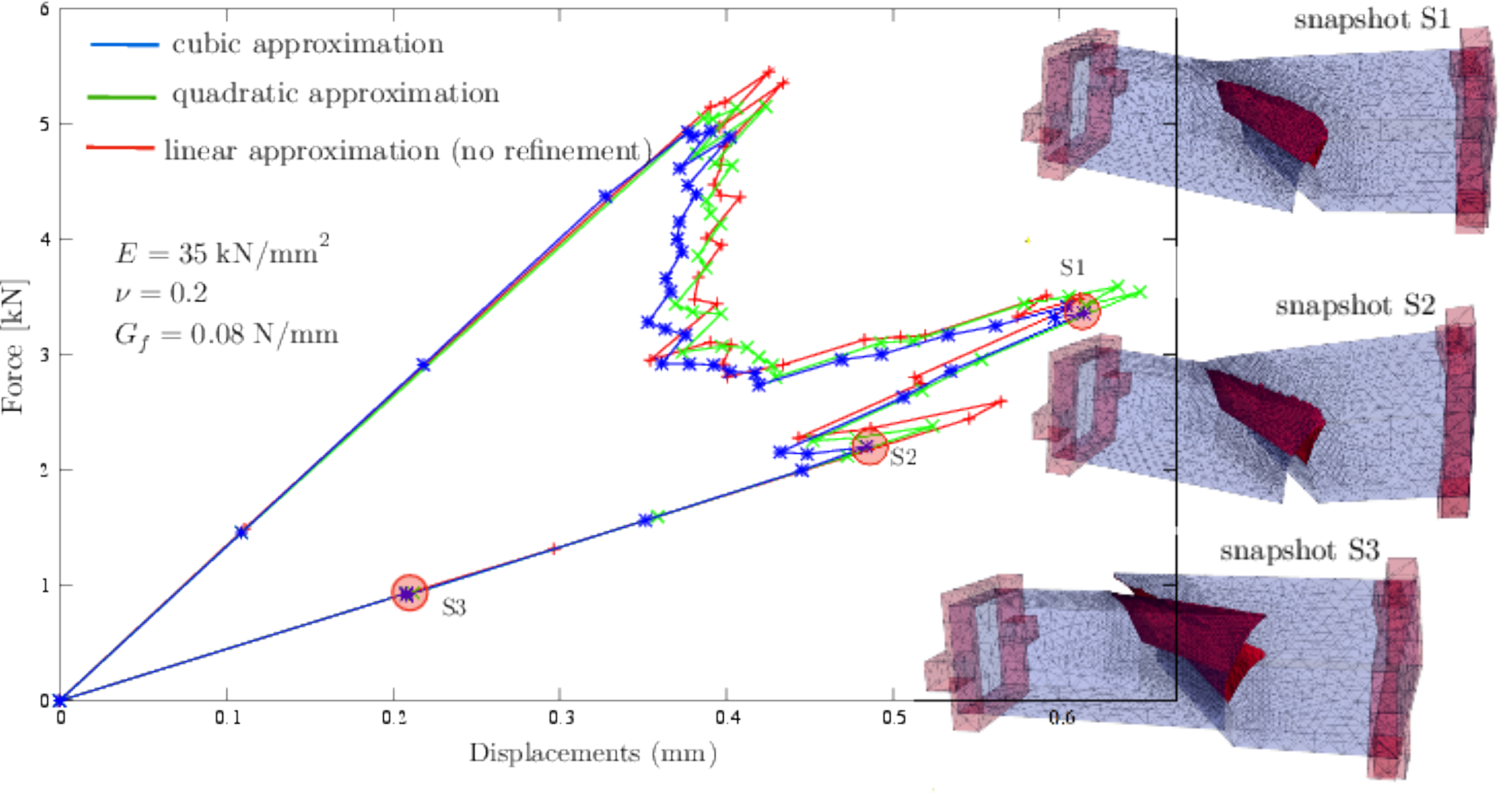}
\end{center}
\caption{Load-displacement path. Animation for this test is
available under this \href{http://youtu.be/FYfDJhFLeXk}{\texttt{link}}.\label{F7}}
\end{figure}
\begin{figure}[htb]
\setlength{\fboxsep}{0pt}%
\setlength{\fboxrule}{0pt}%
\begin{center}
\includegraphics[width=0.9\textwidth]{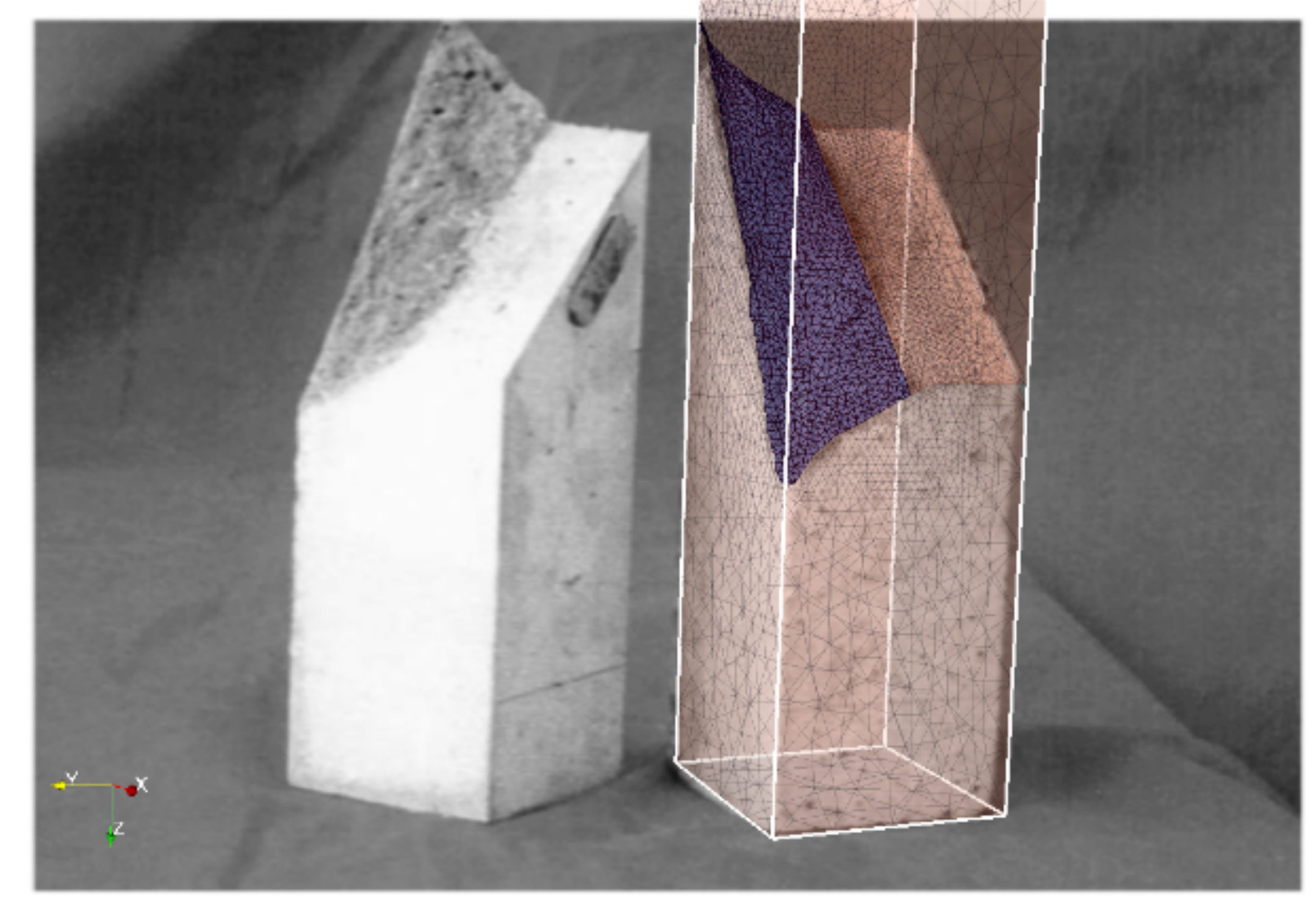}
\end{center}
\caption{The simulated crack surface superimposed on experimental result ~\cite{R22}. \label{F8}}
\end{figure}

\begin{figure}[htb]
\setlength{\fboxsep}{0pt}%
\setlength{\fboxrule}{0pt}%
\begin{center}
\includegraphics[width=0.58\textwidth]{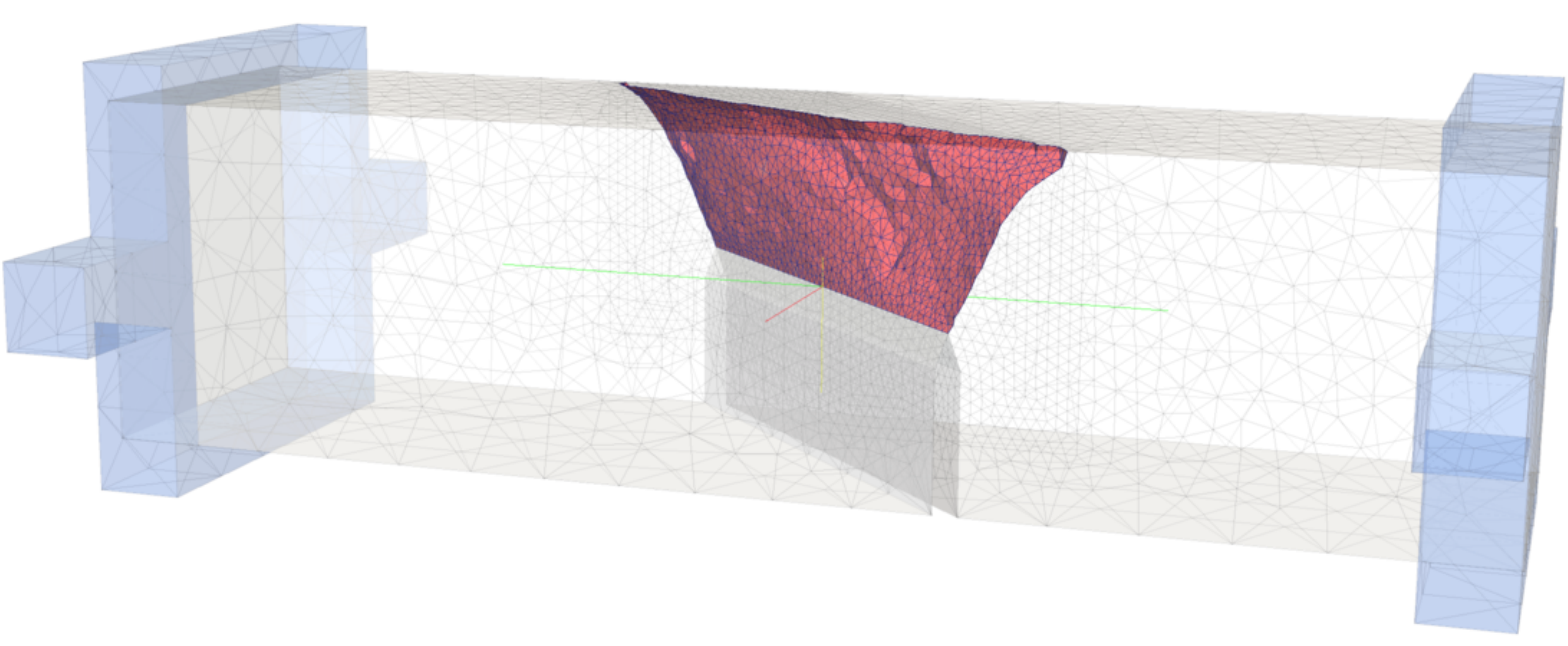}\\
\includegraphics[width=0.58\textwidth]{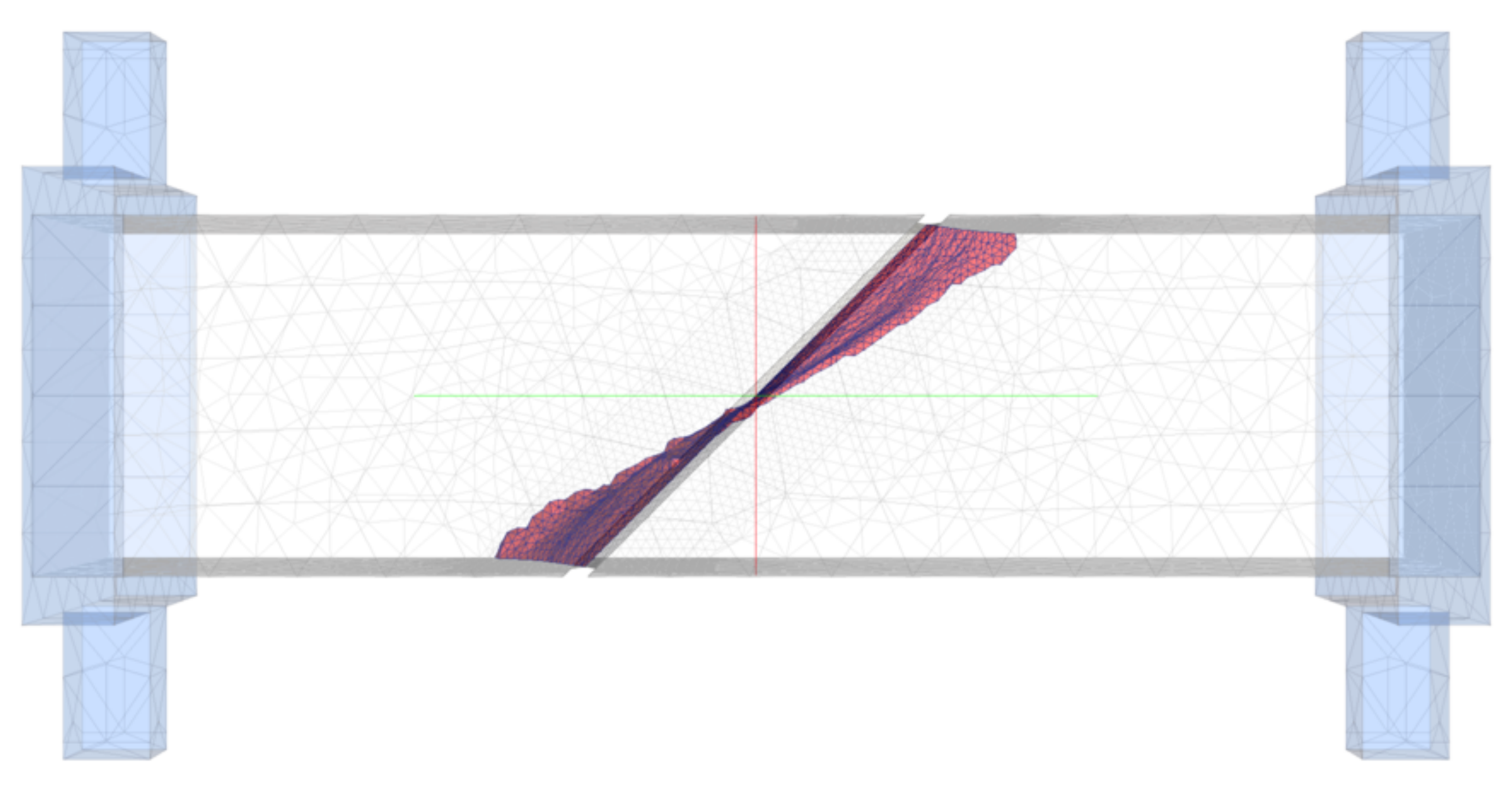}
\end{center}
\caption{The simulated crack surface viewed from the side and above. \label{F9}}
\end{figure}

The beam and steel frame are discretized using tetrahedral
elements. The mesh consists of $29941$ nodes, see Fig.~\ref{F7}. For this problem,
three numerical tests with the same initial mesh are investigated. The first test was
without adaptive mesh refinement near the crack front, the second with one level of mesh
refinement and a quadratic approximation basis near the crack front and
the third with two levels of h-refinement and a cubic hierarchical approximation
basis near the crack front. In order to study
the rate of convergence associated with refinement, the propagating crack is resolved by splitting the
original mesh before refinement (note that in the pull-out test it was the refined mesh that was split). Hence, the mesh refinement
improves the approximation quality of the displacements fields, but the resolution of the crack front geometry is the same for all three tests. 

Fig.~\ref{F7} shows the load-displacement response for the three numerical tests. Good numerical convergence is observed with increasing refinement and the arc-length control method is able to track the dissipative loading path. 
%Smaller values of $\beta$ results in some load steps with
%convergence difficulties and further improvement could be achieved by reducing
%$\alpha_\textrm{min}$ parameter with an adaptive quality control associated with the
%parameter $\alpha$, however this have not been implemented, yet.
Fig.~\ref{F8} shows the numerically predicted geometry of the cracked specimen overlain on to the experimentally observed geometry,
demonstrating excellent agreement.
Despite the good qualitative predictions, the numerical analyses (Fig.~\ref{F7}) over predict the experimental ultimate load by approximately $2.5$ times. This observed difference is a consequence of assuming linear elastic fracture mechanics for a problem where the size of the fracture process zone is significant compared to the size of the problem. This was not an issue in the pull-out test, since the characteristic size of the problem was significantly larger (see Fig.~\ref{F4}). The fracture energy is not only used in the creation of the macroscopic crack surface, but a significant proportion is used to create meso-cracks in the vicinity of the crack front (Ba\v{z}ant~\cite{R22}). Even though the crack geometry matches the experimental results very well, see Fig.~\ref{F8}, the macro-crack has insufficient surface area to match the total amount of dissipated energy. Dissipation of energy in the macro-cracks is not included in the constitutive model and the meso-cracks are not discretised. An enhancement of the constitutive model to include cohesive cracking is an option for future work.

%Fig.~\ref{F9}
%
%
%
%
%
%Note that the Griffith-like theory based on maximal dissipation of energy is
%employed here. 
%
%presented model over predict
%approximately by $2.5$ times the ultimate force measured in the experiment \cite{R22}.
%In spite of the fact there is mismatch between the predicted force and experiment, it is not
%result of errors in the numerical model or an inappropriate numerical
%solution technique, but in violation of underlying assumptions of mechanical model.  It is
%assumed a non existence of characteristic (intrinsic) length of the material,
%whereas in reality the maximum aggregate size is $10$ mm and compared to the
%size of sample this can not be neglected. The reason for such incorrect results is
%well explained by Ba\v zant \cite{R22}, that the fracture energy is not only
%used to the creation of the macroscopic crack surface, but significant part of it is
%used to create a mesocrack in the vicinity of the crack front. This explain the fact that
%presented results over estimate the ultimate strength, since the crack path match
%experimental results well, see Fig.~\ref{F8}, there is no enough macro-crack area to
%match the total amount of dissipated energy, which is significant part of it is
%dissipated to create meso-cracks not directly discretized here nor included
%in the constitutive model.

\section{Conclusions}

This paper has presented a computational model for quasi-static, three-dimensional, brittle fracture based on configurational mechanics. 
The model predicts the direction of crack extension based on the principle of energy minimization in conjunction with a node-based Griffith-like crack criterion. 
The direction of the crack extension is aligned with the nodal configurational forces at the crack front and the crack extension is resolved by locally aligning element faces and then splitting the mesh. Both h- and p-refinement are utilised in the vicinity of the crack front to increase the accuracy of the evolving crack front geometry and the approximation of the spatial displacements. A monolithic solution strategy, simultaneously solving for both the
material displacements (i.e. crack extension) and the spatial displacements, is presented. In order to track the dissipative loading path for the ideally brittle
material, an arc-length procedure has been derived that is tailored for this model.
%It is shown that applying
%the presented methodology, the crack propagation criterion is exclusively expressed
%by the nodal quantities, simply by the out--of--equilibrium residual nodal
%forces in the material space. This provides full consistency between the numerical
%and the continuum mechanical model.

%Moreover in order to track the dissipative loading path for the ideally brittle
%material a problem tailored arc-length control function is derived, which is
%capable to directly control the quasi-static crack surface evolution. 

A key element of the solution strategy is the control of mesh quality for the evolving problem. A pseudo `stress' for the mesh quality measure is derived as
a function of the material gradient of deformation. This mesh quality control is used locally in the vicinity of the crack front following realignment of element faces but it is also used globally to stabilise the Newthon--Raphson method. 
It is worth noting that this mesh smoothing method could be implemented in any standard FE system using the pseudo `stress', standard shape functions and standard procedures for numerical integration and vector and matrix assembly. A method of selecting the tetrahedral element faces for crack
surface approximation is presented. This method is based on an undirected
graph (tree) allowing for efficient and robust search among admissible sets of
the element faces for crack extension, which leads to minimal mesh distortion.

This formulation can be extended to include the case of
anisotropic materials, material heterogeneity and other dissipative processes,
for example to capture nonlinear size effect, when the characteristic size of the
microstructure needs to be included in the model.
To achieve this, thermodynamic constraints on the local state variables can be derived by applying the
Coleman procedure \cite{R7}. This could potentially allow the model to be extended to include dissipative processes related to crack opening, 
applying the cohesive crack methodology \cite{R3,R4}. In such a situation, the energy dissipation associated with a macroscopic crack and dissipation related to meso-level cracking are quantified independently.

The model has been implemented on parallel shared memory computers using libraries such as
PETSc \cite{R18} and MOAB \cite{R19}. The two numerical examples demonstrated the model's ability to accurately simulate complex crack paths and to trace the dissipative load-displacement path. 

\section*{Acknowledgements} This work was supported by EDF Energy Nuclear
Generation Ltd. The views expressed in this paper are those of the authors and not necessarily those of EDF Energy Nuclear Generation Ltd. Analyses were undertaken using the EPSRC funded ARCHIE-WeSt High Performance 
Computer (www.archie-west.ac.uk). EPSRC grant no. EP/K000586/1. We also acknowledge Prof.~Nenad Bi\' cani\'c for his helpful comments.

\end{document}